\documentclass[a4paper,11pt]{article}
\pdfoutput=1 

\usepackage{jheppub} 
\usepackage[T1]{fontenc} 
\usepackage[italian,english]{babel}
\usepackage{hyperref}
\usepackage{ifpdf}
\usepackage{subfigure}
\usepackage{tikz}
\usepackage[compat=1.1.0]{tikz-feynman}
\usepackage{slashed}
\usetikzlibrary{arrows,shapes}
\usetikzlibrary{trees}
\usetikzlibrary{matrix,arrows} 
\usetikzlibrary{positioning}
\usetikzlibrary{calc,through}
\usetikzlibrary{decorations.pathreplacing}  
\usepackage{pgffor}

\usetikzlibrary{decorations.pathmorphing}
\usetikzlibrary{decorations.markings}
\tikzset{
vector/.style={decorate, decoration={snake}, draw},
fermion/.style={draw=black, postaction={decorate}}, 
scalar/.style={dashed,draw=black, postaction={decorate}}}
\tikzstyle{block} = [draw, rectangle, 
minimum height=3em, minimum width=6em]

\usepackage{amssymb}
\usepackage{amsfonts}
\usepackage{epsf}
\usepackage{rotating}
\usepackage{graphicx}
\usepackage{amsmath}
\usepackage{fancyhdr}
\usepackage{lineno}
\usepackage{babel}
\usepackage{graphics}
\usepackage{pstricks}
\usepackage{color}
\usepackage{multirow}
\usepackage{float}
\usepackage{lineno}
\usepackage{mathtools}
\usepackage{stackengine}
\usepackage{comment}

\newcommand{\lsim}{\mathrel{\mathop{\kern 0pt \rlap
{\raise.2ex\hbox{$<$}}}
\lower.9ex\hbox{\kern-.190em $\sim$}}}
\newcommand{\gsim}{\mathrel{\mathop{\kern 0pt \rlap
{\raise.2ex\hbox{$>$}}}
\lower.9ex\hbox{\kern-.190em $\sim$}}}

\newcommand{\be}{\begin{equation}}
\newcommand{\ee}{\end{equation}}
\newcommand{\bea}{\begin{eqnarray}}
\newcommand{\eea}{\end{eqnarray}}

\def\gev{\ensuremath{\mathrm{\,Ge\kern -0.1em V\,}}}
\def\tev{\ensuremath{\mathrm{\,Te\kern -0.1em V\,}}}

\begin{document}

\title{ Neutrino masses and mixed dark matter from doublet and singlet scalars }

\author[a]{Jongkuk Kim}
\author[a]{, Seong-Sik Kim}
\author[a]{, Hyun Min Lee}
\author[a]{, and Rojalin Padhan}

\affiliation[a]{Department of Physics, Chung-Ang University, Seoul 06974, Korea}
\emailAdd{jongkukkim@cau.ac.kr}
\emailAdd{sskim.working@gmail.com}
\emailAdd{hminlee@cau.ac.kr}
\emailAdd{rojalinpadhan2014@gmail.com}


\abstract{We consider the extension of the Standard Model with an inert scalar doublet, three right-handed neutrinos, and singlet scalar fields, $\varphi$ and $S$. In this model, neutrino masses are zero in the limit of the unbroken $Z_4$ discrete symmetry. We show that when the singlet scalar field $\varphi$ gets a VEV, the $Z_4$ symmetry is broken to $Z_2$, and neutrino masses are generated at one-loops due to the mixings between the neutral components of the inert scalar doublet and the singlet scalar field $S$. There is a dark matter candidate from the lightest neutral scalar field, which is a mixture of the inert scalar doublet and the singlet scalar field $S$, in general. The $Z_4$ breaking mass terms are constrained by electroweak precision data and direct detection (DD) bounds for dark matter, favoring  small mixings or almost degenerate masses for the DM scalars. As a result, we discuss the implications of the results for small neutrino masses and DD-safe dark matter. }

\maketitle              

\section{Introduction}

There is no doubt that we need to go beyond the Standard Model due to nonzero masses for neutrinos.
The scotogenic model \cite{Ma:2006km} is among the mechanisms for generating neutrino masses, and it requires an inert scalar doublet (one more scalar doublet with a vanishing VEV) and three right-handed neutrinos. Unlike the high-scale seesaw mechanism \cite{Minkowski:1977sc,Yanagida:1979as,Gell-Mann:1979vob,Mohapatra:1979ia,Schechter:1980gr}, neutrino masses are generated by radiative corrections with the inert scalar doublet and the RH neutrinos running in loops.
So, there is no need to take new particles in the scotogenic model to be much heavier than the weak scale, and the lighter neutral scalar field of the extra Higgs doublet is a good candidate for dark matter (DM). Then, there are testable signatures of the inert scalar doublet by dark matter direct detection, electroweak precision tests and collider experiments.

Recently, there has been a new proposal to extend the scotogenic model with extra singlet scalar fields, $\varphi$ and $S$, where a nonzero VEV of $\varphi$ leads to a small lepton number violating coupling in the scalar sector due to the breakdown of a discrete $Z_4$ symmetry \cite{Kim:2024cwp}.  In this case, each neutral scalar field of the inert doublet can mix with the component field of the complex singlet scalar $S$ with the same CP parity, so small neutrino masses can be ascribed to small dark matter mixings in the limit of the decoupled  field $S$. Thus, the model provides not only an effective description of the lepton number violating quartic coupling $\lambda_{5,{\rm eff}}$ of the original scotogenic model, but also an extra annihilation channel for dark matter into a pair of the dark Higgs scalars.
There is a similar model proposed to explain the neutrino masses based on the $Z_4$ symmetry, but the focus was on the impact of the extra scalar fields on the phenomenology of the RH neutrino dark matter \cite{delaVega:2024tuu}.

In this article, we perform a general phenomenological study of the model with the $Z_4$ symmetry for neutrino masses and scalar dark matter beyond the case where the singlet scalar $S$ is decoupled \cite{Kim:2024cwp}. We propose some benchmark models for obtaining neutrino masses due to the DM mixings and correlate between neutrino masses and dark matter phenomenology in each benchmark model. We consider various possibilities for dark matter in the model: a singlet scalar DM, an inert doublet-like DM, a mixed dark matter from doublet and singlet scalars, etc, and show the importance of the light singlet dark Higgs for dark matter annihilations satisfying the correct relic density while being consistent with vacuum stability, perturbativity, Higgs/electroweak precision data, collider bounds as well as the strong bounds from direct detection experiments.

The paper is organized as follows. We first present the setup for the extension of the SM with extra scalar fields and RH neutrinos and review the motivation of the simple case where small neutrino masses are obtained in the decoupling limit of the singlet scalar $S$. 
Then, we consider the vacuum structure of the model with a broken $Z_4$ symmetry and discuss the Higgs mixing and the mass spectrum and mixings for $Z_2$-odd DM scalar fields. The bounds from vacuum stability and perturbativity are also given.
Next we enumerate various phenomenological constraints on the model, coming from neutrino masses, DM relic density, DM direct detection, electroweak precision data, collider bounds, etc. We continue to apply those constraints on some benchmark models for scalar dark matter, varying the DM mixing angles from small to maximal values.  There are three appendices dealing with dark matter interaction vertices, perturbativity bounds, and loop functions for oblique parameters. Finally, summary and conclusions are drawn.

\section{The setup}

We regard two singlet scalar fields, $S$ and $\varphi$, an inert scalar doublet $H_2$ in addition to the SM Higgs doublet, and three right-handed (RH) neutrinos, $N_{R,i}$ as being transforming under a local $U(1)_X$ as $\phi_i \to e^{i Q_X\theta} \phi_i$ with $\phi_i=\{S, \varphi, H_1, H_2, N_{R,i}\}$ \cite{Kim:2024cwp}. An extra  gauge boson associated with $U(1)_X$ could also be introduced, but we focus the case where the extra gauge boson is decoupled.  On the other hand, we assume that the SM fermions are neutral under the $U(1)_X$ symmetry. Then, the $U(1)_X$ symmetry contains $Z_4$ or $Z_2$ discrete gauge symmetries as subgroups.
There are similar constructions with $Z_4$ symmetry for the extension of the inert doublet models \cite{Belanger:2021lwd,delaVega:2024tuu}, but there is no mass mixing between dark matter candidates in the former case and the lightest RH neutrino is regarded as dark matter in the latter case.

We specify the $U(1)_X$ charge assignments for the extra scalar fields and the  RH neutrinos as $Q_X(\phi_i) = (+1,  +2, 0, +1,+1)$. The $U(1)_X$ charges and $Z_4$ and $Z_2$  parities for scalar fields and the RH neutrinos are summarized in Table 1. 

\begin{table}[!ht]
    \centering
    \begin{tabular}{|c|c|c|c|c|c|}
    \hline
         & $S$ & $\varphi$ & $H_1$ & $H_2$ & $N_{R,i}$ \\
        \hline\hline
        $U(1)_X$ & $+1$ & $+2$ & $0$ & $+1$ & $+1$  \\
        \hline
        $U(1)_X\supset Z_4$  &  $i$  &  $-1$ & $1$ & $i$ & $i$ \\
        \hline 
        $U(1)_X\supset Z_2$   &   $-1$    &  $+1$ & $+1$ & $-1$ & $-1$ \\
        \hline
    \end{tabular}
    \caption{The $U(1)_X$ charges, $Z_4$ and $Z_2$ parities. }
    \label{table:1}
\end{table}

We consider the scalar potential in our model, which is composed of $V=V_1+V_2$.
First, the scalar potential for two Higgs doublets only is given by
\begin{align}
V_1 &= m^2_1 \vert H_1 \vert^2 + m^2_2 \vert H_2 \vert^2 +\lambda_1 \vert H_1 \vert^4 +\lambda_2 \vert H_2 \vert^4 + \lambda_3 \vert H_1 \vert^2 \vert H_2 \vert^2 + \lambda_4 ( H^\dagger_1 H_2 )( H^\dagger_2 H_1).
\end{align}
Here, we note that the $\lambda_5$ term appearing in the inert double dark matter (IDM) models \cite{Arcadi:2024ukq,Gong:2012ri} is forbidden because of $U(1)_X$ or $Z_4$ symmetries, 
\bea
V_1\supset \frac{\lambda_5}{2} \Big( (H^\dagger_1 H_2)^2 + (H^\dagger_2 H_1)^2 \Big)=0.
\eea
The part of the scalar potential containing the singlet scalar fields respecting the $U(1)_X$ symmetry is
\begin{align}
V_2&=  \sqrt{2}\kappa S^\dagger H^\dagger_1 H_2+ \lambda'_{S\varphi} \varphi^\dagger  S H^\dagger_1 H_2 + \sqrt{2}\mu \, \varphi^\dagger S^2+ {\rm h.c.} \nonumber  \\
& \quad+m^2_S \vert S\vert^2+\lambda_S \vert S\vert^4 +m^2_\varphi \vert \varphi \vert^2 +\lambda_\varphi \vert \varphi \vert^4 \nonumber\\
&\quad  + \sum_{i=1,2}\bigg[\lambda_{H_i S}\left(H^\dagger_i H_i \right) \vert S \vert^2  + \lambda_{H_i \varphi}\left(H^\dagger_i H_i \right) \vert \varphi \vert^2\bigg]+ \lambda_{S\varphi} \vert S \vert^2 \vert \varphi \vert^2.
\end{align}
Here, we note that if the $Z_4$ symmetry is imposed instead of the full $U(1)_X$, we can also include the extra terms for the singlet scalar for $\varphi$, such as $\varphi S^2, \varphi^2, \varphi^4$, $\varphi^2 |H_1|^2$, $\varphi^2 |H_2|^2$, etc. Even in this case, after $Z_4$ is broken to $Z_2$ due to the VEV of $\varphi$, the effective $Z_2$ models share the same feature as in the $U(1)_X$ case that there is a possibility for the mixing between doublet scalars and singlet scalars of $S$. 

We also introduce the $U(1)_X$-invariant Yukawa couplings for the inert doublet and the RH neutrinos in our model, as follows,
\bea
{\cal L}_{\rm Yukawa}= -y_{N,ij} {\bar l}_i {\widetilde H}_2 N_{R,j} -\frac{1}{2}\lambda_{N,ij}\varphi \overline{N^c_{R,i}} N_{R,j} +{\rm h.c.}  \label{Yukawa}
 \eea
 with $ {\widetilde H}_2=i\sigma_2 H^*_2$. 

We assume that neither $H_2$ or $S$ have a VEV for positive $m^2_2$ and  $m^2_S$,
while $H_1$ and $\varphi$ have non-zero VEVs. 
Then, the VEV of $\varphi$, namely, $\langle\varphi\rangle=\frac{1}{\sqrt{2}} v_\varphi$, breaks $Z_4$ to $Z_2$, leading to the effective $Z_2$ models for dark matter. Alternatively, we can regard $\varphi$ as a dark composite field breaking the $Z_4$ symmetry explicitly due to non-perturbative effects \cite{Lee:2010gv,Lee:2011dya}.

Below the $U(1)_X$ symmetry breaking scale, we obtain the effective potential only for $S, H_1, H_2$ in  $V_2$ where the $Z_4$ symmetry is also spontaneously broken to $Z_2$, as follows,
\bea
V_{2,{\rm eff}} &=& \sqrt{2}\kappa S^\dagger H^\dagger_1 H_2+\sqrt{2}\kappa' S H^\dagger_1 H_2+\frac{1}{2}{\hat m}^2_S S^2 +{\rm h.c.} \nonumber \\
&&+m^2_S \vert S\vert^2+\lambda_S \vert S\vert^4 + \sum_{i=1,2}\lambda_{H_i S}\left(H^\dagger_i H_i \right) \vert S \vert^2, \label{Veff:Z2}
\eea
where $\kappa'=\frac{1}{2}\lambda'_{S\varphi} v_\varphi$  and ${\hat m}^2_S=2 \mu v_\varphi$.  Moreover, the mass terms for the RH neutrinos are generated by ${\cal L}_{\rm RH}=-\frac{1}{2}M_{N,ij} \overline{N^c_{R,i}} N_{R,i} +{\rm h.c.}$, with $M_{N,i}=\frac{1}{\sqrt{2}}\lambda_{N,ij}v_\varphi$.
Here, the mass for the singlet $S$ is shifted by the contribution coming from $\lambda_{S\varphi} v^2_\varphi$, but we keep the notation for $m^2_S$ the same.  The effective $Z_2$ models including the singlet scalar $S$ leads to a flexibility for dark matter due to the mixing between $S$ and the neutral components of $H_2$, as will be discussed in the later section. 

If we further integrate out the singlet scalar $S$ with ${\hat m}^2_S=\sqrt{2} \mu v_\varphi\ll m^2_S$ and $\kappa'\ll m_S$, from the combination of the $Z_2$ invariant terms, $\sqrt{2}\kappa S^\dagger H^\dagger_1 H_2, \sqrt{2}\kappa' S H^\dagger_1 H_2$ and $\frac{1}{2}{\hat m}^2_S S^2$, in eq.~(\ref{Veff:Z2}), we obtain the inert dark matter model with the effective $\lambda_5$ term \cite{Kim:2024cwp} by
\bea
\lambda_{5,{\rm eff}}=\frac{2{\hat m}^2_S (\kappa^2+\kappa^{\prime 2})}{m^4_S}-\frac{4\kappa\kappa'}{m^2_S}. \label{lambdaeff}
\eea
As a result, together with the Yukawa interactions of the inert doublet to the RH neutrinos in eq.~(\ref{Yukawa}), we can obtain the neutrino mass matrix for $|\lambda_{5,{\rm eff}}|\ll 1$ at loop order \cite{Ma:2006km,Kim:2024cwp} as
\bea
({\cal M}_\nu)_{ij} \simeq \frac{\lambda_{5,{\rm eff}}v^2_H}{16\pi^2}\sum_k \frac{y_{N,ik}y_{N,jk}M_{N,k}}{m^2_0-M^2_{N,k}}\bigg[1-\frac{M^2_{N,k}}{m^2_0-M^2_{N,k}}\ln\frac{m^2_0}{M^2_{N,k}}\bigg],
\eea
with $m^2_0\simeq m^2_2+\frac{1}{2}(\lambda_3+\lambda_4)v^2_H$.

However, for the general neutrino and DM phenomenology in our model, we will  include the singlet scalars, $\varphi$ and $S$, explicitly in the later discussion.

\section{Vacuum structure and scalar masses}

We first discuss the minimization of the potential and the vacuum structure of the model and consider the Higgs-singlet mixing and the DM scalar mixings. We also present the vacuum stability and perturbativity bounds in our model.

\subsection{Minimization of the scalar potential}

Taking the unitarity gauge for the SM Higgs, we expand two Higgs doublets as
\begin{align}
H_1 = \frac{1}{\sqrt{2} } \begin{pmatrix} 0  \\ h+v_H 
\end{pmatrix},~~~~  H_2 &= \begin{pmatrix}
H^+ \\ \frac{1}{\sqrt{2}} \left(H_0 + i A_0 \right)
\end{pmatrix}.
\end{align}
On the other hand, the $SU(2)$-singlet scalar fields, $S$ and $\varphi$ are expanded as
\begin{align}
\varphi &= \frac{1}{\sqrt{2}} \left( \rho + v_\varphi +i\eta \right), ~~ S= \frac{1}{\sqrt{2}} \left( s + i a \right).
\end{align}
For the local $U(1)_X$, $\eta$ corresponds to the would-be Goldstone boson, which is set to zero in unitary gauge. In this case, we can regard the gauge boson $X$ associated with $U(1)_X$ as being decoupled due to its mass  $m_X= 2 g_X v_\varphi$, with $g_X$ being the extra gauge coupling, so we can focus on the dynamics of the dark Higgs in the $U(1)_X$ symmetry breaking sector at low energy. Alternatively, without introducing an extra gauge boson, we can consider the $Z_4$ invariant model with a real singlet scalar, namely, $\varphi=\frac{1}{\sqrt{2}}(\rho+v_\varphi)$, which transforms as $\varphi\to -\varphi$ under $Z_4$ according to Table~\ref{table:1}, although the DM mass spectrum and the DM self-interactions also depend on the extra $Z_4$-invariant terms for $\varphi$ in the potential, such as $\varphi S^2, \varphi^2, \varphi^4$, $\varphi^2 |H_1|^2$, $\varphi^2 |H_2|^2$, etc.  In both cases with a local $U(1)_X$ and a discrete $Z_4$ symmetry, the essence of the low-energy correlation between dark matter and  neutrino masses can be retained.

On the other hand, if $U(1)_X$ is a global symmetry, it is broken spontaneously by the VEV of $\varphi$, so $\eta$ would be a massless Goldstone boson, which is Majoron-like as it couples to the RH neutrinos. However, there is no mass mixing between the active neutrinos and the RH neutrinos in our model, so there is no direct Majoron coupling to the active neutrinos. Nonetheless, if $\eta$ is a pseudo-Goldstone boson, it could make sizable contributions to dark radiation and dark matter annihilations for a small $v_\varphi$, and spontaneous leptogenesis might be possible as an alternative to thermal leptogenesis \cite{Kim:2024cwp}. However, we postpone the case with a global $U(1)_X$ as another work.

Now we identify the minimum of the potential in our model. To this, we rewrite the potential in terms of the CP-even scalars,  $\chi_i=h, \rho, H_0, s$, as follows,
\bea
V&=&\frac{1}{2} m^2_1 h^2 + \frac{1}{2} m^2_2 (H_0)^2 + \frac{1}{4} \lambda_1 h^4 + \frac{1}{4} \lambda_2 (H_0)^4 +\frac{1}{4} (\lambda_3+\lambda_4) h^2 (H_0)^2 \nonumber \\
&&+\kappa s H_0 h + \frac{1}{2} \lambda'_{S\varphi} \rho h s H_0 +\mu \rho s^2 + \frac{1}{2} m^2_S s^2 +\frac{1}{4} \lambda_S s^4 + \frac{1}{2} m^2_\varphi \rho^2 + \frac{1}{4} \lambda_\varphi \rho^4 \nonumber \\
&&+ \frac{1}{4} (\lambda_{H_1 S}h^2+ \lambda_{H_2 S} (H_0)^2) s^2 + \frac{1}{4} (\lambda_{H_1\varphi} h^2+ \lambda_{H_2\varphi} (H_0)^2 )\rho^2+\frac{1}{4}\lambda_{S\varphi} s^2 \rho^2. 
\eea
Here, we can determine the VEVs by imposing the first derivatives of the CP-even scalar fields to vanish, i.e. $\frac{\partial V}{\partial \chi_i}$=0,  with
\begin{align}
\frac{\partial V}{\partial h} &= m^2_1h +\lambda_1 h^3  + \frac{1}{2}(\lambda_3+\lambda_4)h (H_0)^2+\kappa sH_0  \nonumber\\
&\quad+\frac{1}{2}\lambda'_{S\varphi}\rho s H_0+ \frac{1}{2} \lambda_{H_1 S} h s^2 + \frac{1}{2} \lambda_{H_1 \varphi}h \rho^2,\\
\frac{\partial V}{\partial \rho} &= m^2_\varphi \rho +\lambda_\varphi \rho^3 +\frac{1}{2}\lambda'_{S\varphi} h s H_0 + \mu s^2  \nonumber\\
&\quad+ \frac{1}{2}\lambda_{H_1 \varphi}h^2\rho + \frac{1}{2}\lambda_{H_2 \varphi}(H_0)^2\rho  +\frac{1}{2} \lambda_{S\varphi} s^2\rho , \\
\frac{\partial V}{\partial H_0} &= m^2_2 H_0 +\lambda_2 (H_0)^3 +\frac{1}{2}(\lambda_3+\lambda_4)h^2 H_0 +\kappa s h \nonumber \\
&\quad + \frac{1}{2}\lambda'_{S\varphi} \rho h s +\frac{1}{2}\lambda_{H_2S} H_0 s^2 + \frac{1}{2}\lambda_{H_2\varphi} H_0 \rho^2,   \\  
\frac{\partial V}{\partial s} &= m^2_S s+\lambda_S s^3 +\kappa H_0 h + \frac{1}{2}\lambda'_{S\varphi} \rho h H_0 +2 \mu \rho s\nonumber \\
&\quad + \frac{1}{2} \lambda_{H_1S} h^2 s + \frac{1}{2} \lambda_{H_2S} (H_0)^2 s + \frac{1}{2}\lambda_{S\varphi}s\rho^2.
\end{align}

Then, we  take the VEVs of $h$ and $\rho$ fields to $v_H$ and $v_\varphi$, respectively, whereas setting the VEVs of the other fields  to zero.
As a result, the minimization conditions for the potential gives rise to
\begin{align}
m^2_1 &= -\lambda_1 v_H^2 - \frac{1}{2}\lambda_{H_1 \varphi} v^2_\varphi , \\
m^2_\varphi &=  - \lambda_{\varphi} v^2_\varphi - \frac{1}{2}\lambda_{H_1 \varphi} v^2_H.
\end{align}
Therefore, we can determine the VEVs as
\bea
v_H^2 &=& 2\,\frac{\lambda_{H_1 \varphi} m^2_\varphi-2\lambda_\varphi m^2_1 }{4\lambda_1\lambda_\varphi-\lambda_{H_1 \varphi}^2}, \label{vH} \\
v_\varphi^2 &=&2\,\frac{\lambda_{H_1 \varphi} m^2_1-2\lambda_1 m^2_\varphi }{4\lambda_1\lambda_\varphi-\lambda_{H_1 \varphi}^2}. \label{vphi}
\eea

Moreover, we also get the mass matrix for $\{h, \rho\}$ as
\bea
{\cal M}^2_{h,\rho}=\left(\begin{array}{cc}m^2_1+3\lambda_1 v^2_H +\frac{1}{2}\lambda_{H_1 \varphi} v^2_\varphi &\lambda_{H_1\varphi} v_H v_\varphi    \\ \lambda_{H_1\varphi} v_H v_\varphi & m^2_\varphi+3\lambda_\varphi v^2_\varphi +\frac{1}{2}\lambda_{H_1\varphi} v^2_H  \end{array}\right), \label{HiggsM}
\eea 
and the mass matrix for $\{H_0, s\}$  as
\bea
{\cal M}^2_{H_0,s}=\left(\begin{array}{cc}m^2_2+\frac{1}{2}(\lambda_3+\lambda_4)v^2_H+\frac{1}{2}\lambda_{H_2\varphi} v^2_\varphi  & \kappa v_H+ \frac{1}{2}\lambda'_{S\varphi} v_H v_\varphi \\   \kappa v_H+ \frac{1}{2}\lambda'_{S\varphi} v_H v_\varphi &  m^2_S +2\mu v_\varphi+\frac{1}{2} \lambda_{H_1S} v^2_H +\frac{1}{2} \lambda_{S\varphi} v^2_\varphi \end{array}\right). \label{DMM}
\eea
Here, we note that there is no mixing between $\{h, \rho\}$ and  $\{H_0, s\}$. 
Thus, the vacuum with $v_H\neq 0$ and $v_\varphi\neq 0$ is a stable local minimum, as far as the Hessians are positive, namely, ${\rm det}({\cal M}^2_{h,\rho})>0$ and ${\rm det}({\cal M}^2_{H_0,s})>0$. 

By using the results in eqs.~(\ref{vH}) and (\ref{vphi}), the diagonal elements of ${\cal M}^2_{h,\rho}$ become $2\lambda_1 v^2_H$ and $2\lambda_\varphi v^2_\varphi$, respectively, so  ${\rm det}({\cal M}^2_{h,\rho})=(4\lambda_1\lambda_\varphi-\lambda_{H_1 \varphi}^2) v^2_H v^2_\varphi>0$.
Thus, together with eqs.~(\ref{vH}) and (\ref{vphi}), we find the conditions for a local minimum \cite{Lebedev:2011aq} as
\bea
&&4\lambda_1\lambda_\varphi-\lambda_{H_1 \varphi}^2>0, \\
&& \lambda_{H_1 \varphi} m^2_\varphi-2\lambda_\varphi m^2_1>0, \\
&&\lambda_{H_1 \varphi} m^2_1-2\lambda_\varphi m^2_\varphi>0. 
\eea

\subsection{Higgs mixing}

The SM Higgs mixes with the neutral component of singlet scalar $\varphi$ from the mass matrix in eq.~(\ref{HiggsM}), which is rewritten as
\begin{align}
{\cal M}^2_{h, \rho}=\begin{pmatrix} m^2_{h} &  \lambda_{H_1\varphi} v_H v_\varphi \\  \lambda_{H_1\varphi}  v_H v_\varphi& m^2_\varphi   \end{pmatrix}
\end{align}
where $m^2_h =2\lambda_1 v^2_H$ and $m^2_\varphi=2\lambda_\varphi v^2_\varphi$.
Then, we can diagonalize the mass matrix by introducing the mixing matrices, as follows,
\begin{align}
	\begin{pmatrix}  h \\ \rho \end{pmatrix} =  \begin{pmatrix}  \cos \alpha & -\sin \alpha \\ \sin \alpha & \cos \alpha \end{pmatrix} \begin{pmatrix} h_1 \\ h_2 \end{pmatrix}
\end{align}
with
\begin{align}
\tan 2\alpha &= {2v_H v_\varphi \lambda_{H_1\varphi} \over m^2_{h}- m^2_\varphi }.
\end{align} 
Then, the mass eigenvalues for neutral scalars are given by
\bea
m^2_{h_1, h_2} &=&\frac{1}{2}\bigg[m^2_{h}+m^2_\varphi \pm (m^2_{h}-m^2_\varphi) \sqrt{1+\frac{4 v_H^2 v_\varphi^2 \lambda_{H_1\varphi}^2 }{(m^2_{h}-m^2_\varphi)^2 }} \bigg].
\eea
We note that if $\lambda_{H_1\varphi}$ is sizable, the Higgs mixing gets suppressed for $v_\varphi \gg v_H$ to be consistent with Higgs data. But, for $v_\varphi\sim v_H$, we can take a small $\lambda_{H_1\varphi}$.

\subsection{Dark matter mixing}

Before considering the mixing mass terms, we first list the scalar masses for the inert Higgs doublet and the real scalars, $s, a$, of the singlet $S$ as
\begin{align}
m^2_{H^\pm} &= m^2_2 +\frac{1}{2 }\lambda_3 v^2_H+ \frac{1}{2}\lambda_{H_2\varphi} v^2_\varphi, \\
m^2_{H_0} &=  m^2_2 + \frac{1}{2}\lambda_Lv^2_H +\frac{1}{2} \lambda_{H_2\varphi} v^2_\varphi =m^2_{A_0}, \\
m^2_s&=m^2_S +\frac{1}{2}\lambda_{H_1S} v^2_H +{\hat m}^2_S +\frac{1}{2} \lambda_{S\varphi} v^2_\varphi =m^2_a+2{\hat m}^2_S,
\end{align}
with $\lambda_L\equiv \lambda_3+\lambda_4$ and ${\hat m}^2_S=2 \mu v_\varphi$.
Due to the $Z_4$ symmetry, we set $\lambda_5=0$, so the masses of $H_0$ and $A_0$ fields are the same, that is, $m^2_{H_0}=m^2_{A_0}$.
We note that a negative $\lambda_4$ is taken for the neutral Higgs to be lighter than the charged Higgs in IDM.

In our model, however, there are mass mixings between the neutral scalars of the inert doublet and the singlet scalars containing in $S$, leading to split masses between neutral scalar fields.
First, in the basis of CP-even scalars, $(H_0, s)$, we rewrite the mass matrix in eq.~(\ref{DMM}) as
\begin{align}
{\cal M}^2_{H_0, s}=\begin{pmatrix} m^2_{H_0} & (\kappa+\kappa') v_H \\ (\kappa+\kappa') v_H  & m^2_s   \end{pmatrix},
\end{align}
with $\kappa'=\frac{1}{2}\lambda'_{S\varphi} v_\varphi$. 
We also obtain the mass matrix  for the CP-odd scalars, $(A_0, a)$,  as
\begin{align}
{\cal M}^2_{A_0, a}=\begin{pmatrix} m^2_{H_0} & (\kappa-\kappa') v_H \\ (\kappa-\kappa') v_H  & m^2_a  \end{pmatrix}.
\end{align}
Then, we can diagonalize the mass matrices by introducing the mixing matrices, as follows,
\begin{align}
	\begin{pmatrix}  H_0 \\ s \end{pmatrix} =  \begin{pmatrix}  \cos\theta_s  & -\sin\theta_s \\ \sin\theta_s & \cos\theta_s \end{pmatrix} \begin{pmatrix} H_1 \\ H_2 \end{pmatrix}, \nonumber\\
	\begin{pmatrix}  A_0 \\ a \end{pmatrix} =  \begin{pmatrix}  \cos\theta_a  & -\sin\theta_a\\ \sin\theta_a & \cos\theta_a \end{pmatrix} \begin{pmatrix} A_1 \\ A_2 \end{pmatrix},
\end{align}
with
\begin{align}
\tan 2\theta_s &=  {2(\kappa+\kappa') v_H \over m^2_{H_0}-m^2_s }, \\
\tan 2\theta_a &= {2(\kappa-\kappa') v_H \over m^2_{H_0}- m^2_a }.
\end{align} 
Then, the mass eigenvalues for neutral scalars are given by
\bea
m^2_{H_{1,2}} &=&\frac{1}{2}\bigg[m^2_{H_0}+m^2_s\pm (m^2_{H_0}-m^2_s) \sqrt{1+\frac{4(\kappa+\kappa')^2 v^2_H}{(m^2_{H_0}-m^2_s)^2 }} \bigg],  \label{Hmasses} \\
m^2_{A_{1,2}} &=&\frac{1}{2}\bigg[m^2_{H_0}+m^2_a\pm (m^2_{H_0}-m^2_a) \sqrt{1+\frac{4(\kappa-\kappa')^2 v^2_H}{(m^2_{H_0}-m^2_a)^2 }} \bigg].
\label{Amasses}
\eea
We note that we used the notations for the mass eigenstates of the CP-even neutral scalars by $H_{1,2}$, which are the same as for the Higgs doublets, but we remind ourselves of the notations whenever necessary.

Therefore, the lightest neutral scalar among $H_{1,2}$ and $A_{1,2}$ becomes a candidate for dark matter. In this case, even if $m^2_{H_0}=m^2_{A_0}$ due to the $Z_4$ symmetry, there arises a mass splitting between the CP-even and CP-odd scalars due to $m^2_s\neq m^2_a$ and $\kappa'\neq 0$, which stem from the $Z_4$ breaking mass terms for the singlet scalar. So, when the mass splitting is larger than about $ 100$ keV, we can avoid the strong bound from direct detection for the inelastic DM-nucleon scattering mediated by $Z$ boson.

\subsubsection{Small DM mixings}

We first consider the limit with small mixings for dark scalars.
Taking $|\kappa+\kappa'| \ll |m^2_{H_0}-m^2_s|/(2v_H)$ and $|\kappa-\kappa'| \ll |m^2_{H_0}-m^2_a|/(2v_H)$, we can approximate the mass eigenvalues in eqs.~(\ref{Hmasses}) and (\ref{Amasses}) as
\bea
m^2_{H_1}&\simeq& m^2_{H_0}+ {\rm sgn}(m^2_{H_0}-m^2_s)\,\cdot\frac{(\kappa+\kappa')^2 v^2_H}{|m^2_{H_0}-m^2_s|}, \\
m^2_{H_2}&\simeq& m^2_{s}- {\rm sgn}(m^2_{H_0}-m^2_s)\,\cdot\frac{(\kappa+\kappa')^2v^2_H}{|m^2_{H_0}-m^2_s|}, \\
m^2_{A_1}&\simeq& m^2_{H_0}+ {\rm sgn}(m^2_{H_0}-m^2_a)\,\cdot\frac{(\kappa-\kappa')^2 v^2_H}{|m^2_{H_0}-m^2_a|}, \\
m^2_{A_2}&\simeq& m^2_{a}- {\rm sgn}(m^2_{H_0}-m^2_a)\, \cdot \frac{(\kappa-\kappa')^2 v^2_H}{|m^2_{H_0}-m^2_a|}.
\eea

First,  if $m_{H_0}<m_s, m_a$, the lighter component of the inert doublet-like scalar becomes a dark matter candidate, with the mass splitting,
\bea
m^2_{H_1}-m^2_{A_1} &\simeq & \frac{(\kappa+\kappa')^2 v^2_H}{m^2_{H_0}-m^2_s}-\frac{(\kappa-\kappa')^2 v^2_H}{m^2_{H_0}-m^2_a} \nonumber  \\ 
&=&\frac{2(\kappa^2+\kappa^{\prime 2}){\hat m}^2_S v^2_H + 4\kappa\kappa' (m^2_{H_0}-m^2_{s,0})v^2_H}{(m^2_{H_0}-m^2_s)(m^2_{H_0}-m^2_a)}.
\eea
Here, $m^2_{s,0}\equiv m^2_s-{\hat m}^2_S$.
Thus, $A_1$ is always a dark matter candidate for $ {\hat m}^2_S>0$ and $\kappa\kappa'<0$. But, for $\kappa\kappa'>0$,  there is a possibility that $H_1$ is a dark matter candidate. We note that for $m_S\gg m_{H_0}, |{\hat m}_S|$, for which $m_s, m_a\simeq m_S$, the above result becomes
\bea
m^2_{H_1}-m^2_{A_1} \simeq \frac{2(\kappa^2+\kappa^{\prime 2}){\hat m}^2_S v^2_H}{m^4_S} -\frac{4\kappa\kappa' v^2_H}{m^2_S}=\lambda_{5,{\rm eff}} \, v^2_H, \label{eq:massdeiff}
\eea
which is consistent with the effective $\lambda_5$ term in eq.~(\ref{lambdaeff}), obtained in the decoupling limit of the singlet scalars.
For a small mass splitting for the inert doublet-like scalars, $H_1$ decays into $A_1$ and a pair of neutrinos, mediated by the $Z$ boson, so the corresponding decay rate is $\Gamma\propto \Delta^5$ with $\Delta = m_{H_1}-m_{A_1}$.

Next, we consider the case with $m_{H_0}>m_s, m_a$.  Then, the lighter component of the singlet-like scalars is a dark matter candidate: $A_2$ for  $ {\hat m}^2_S>0$ and $H_2$ for  $ {\hat m}^2_S<0$, with the mass splitting given by
\bea
m^2_{H_2}-m^2_{A_2} \simeq m^2_s-m^2_a= 2 {\hat m}^2_S.
\eea

Lastly, if $m_s>m_{H_0}>m_a$ or $m_a>m_{H_0}>m_s$, the lighter component of the singlet-like scalars is still a dark matter candidate: $A_2$ in the former case and $H_2$ in the latter case, with the mass splittings given by
\bea
m^2_{H_1}-m^2_{A_2} &\simeq& m^2_{H_0}-m^2_a,\\
m^2_{A_1}-m^2_{H_2} &\simeq& m^2_{H_0}-m^2_s.
\eea

\subsubsection{Maximal DM mixings}

We also consider the limit with maximal mixings for dark scalars.
If $\theta_s=\theta_a=\frac{\pi}{4}$,  we need $m_{H_0}=m_s=m_a$, so the mass eigenvalues become
\bea
m^2_{H_{1,2}}&=& m^2_{H_0}\pm |\kappa+\kappa'|v_H, \\
m^2_{A_{1,2}}&=&m^2_{H_0}\pm |\kappa-\kappa'|v_H.
\eea
Then, we find that $A_2$ is the lightest neutral scalar for $\kappa\kappa'<0$ while $H_2$  the lightest neutral scalar for $\kappa\kappa'>0$.
 
If $\theta_s=\frac{\pi}{4}$ but $|\theta_a|\ll 1$, we obtain $m_{H_0}=m_s$ and
\bea
m^2_{H_{1,2}}&=&m^2_{H_0}\pm |\kappa+\kappa'|v_H, \\ 
m^2_{A_1}&\simeq& m^2_{H_0}+ {\rm sgn}(m^2_s-m^2_a)\,\cdot\frac{(\kappa-\kappa')^2 v^2_H}{|m^2_s-m^2_a|}, \\
m^2_{A_2}&\simeq& m^2_{a}- {\rm sgn}(m^2_s-m^2_a)\, \cdot \frac{(\kappa-\kappa')^2 v^2_H}{|m^2_s-m^2_a|}.
\eea
This is the case for $|\kappa-\kappa'| v_H\ll |m^2_s-m^2_a|/2$.  Then, unless $\kappa,\kappa'$ are tuned to each other, the mass splitting between $H_1$ and $H_2$ is small too.  
As a result, we obtain the hierarchy of the DM scalars as $m_{H_1}>  m_{A_1}> m_{H_2}>m_{A_2}$ for $m_s>m_a$ and $m_{A_2}>m_{H_1}>  m_{A_1}> m_{H_2}$ for $m_s<m_a$.
Singlet-like scalars, $A_2$ or $H_2$, become a dark matter candidate in the former or latter cases, respectively.

Similarly, if $\theta_a=\frac{\pi}{4}$ but $|\theta_s|\ll 1$, we obtain $m_{H_0}=m_a$ and
\bea
m^2_{A_{1,2}}&=&m^2_{H_0}\pm |\kappa-\kappa'|v_H, \\ 
m^2_{H_1}&\simeq& m^2_{H_0}+ {\rm sgn}(m^2_a-m^2_s)\,\cdot\frac{(\kappa+\kappa')^2 v^2_H}{|m^2_s-m^2_a|}, \\
m^2_{H_2}&\simeq& m^2_{s}- {\rm sgn}(m^2_a-m^2_s)\,\cdot\frac{(\kappa+\kappa')^2v^2_H}{|m^2_s-m^2_a|}.
\eea
This is the case for  $|\kappa+\kappa'|v_H \ll |m^2_s-m^2_a|/2$. Then, unless $\kappa,\kappa'$ are tuned to each other, the mass splitting between $A_1$ and $A_2$ is small too.  
As a result, we obtain the hierarchy of the DM scalars as $m_{A_1}>  m_{H_1}> m_{A_2}>m_{H_2}$ for $m_s<m_a$ and $m_{H_2}>m_{A_1}>m_{H_1}>  m_{A_2}$ for $m_s>m_a$.
Singlet-like scalars, $H_2$ or $A_2$, become a dark matter candidate in the former or latter cases, respectively.

\subsection{Vacuum stability}

To ensure the vacuum stability of the scalar potential, we need to check if it is bounded from below, i.e. $V\ge 0$ in the limit of large field values. Then, it is sufficient to take the quartic part of the scalar potential, $V_4$ and  demand $V_4\ge 0$.

In order to derive the vacuum stability conditions, it is convenient to write $V_4$ in terms of  the squares of real fields as $V_4=X_{ij} r_i^2 r_j^2$. Then, co-positivity of the matrix $X$ ensures the vacuum stability.
We parameterise the bilinears as $|H_1|^2=r^2 c_\beta^2$, $|H_2|^2=r^2 s_\beta^2$, $H_1^\dagger H_2=\chi r^2 c_\beta s_\beta  e^{i\Sigma_1}$, $|S|^2=\rho^2 c_\alpha^2$, $|\varphi|^2=\rho^2 s_\alpha^2$ and$\varphi^\dagger S=\zeta \rho^2 c_\alpha s_\alpha  e^{i\Sigma_2}$. Here, $r,\rho=[0,\infty]$, $\alpha,\beta=[0,\pi/2]$ and $|\chi|,|\zeta|=[0,1]$. With this parameterization, $V_4$ takes the following form,
\begin{align}
V_4&=r^4(\lambda _1  c_{\beta }^4+\lambda _2  s_{\beta }^4+\lambda _3 c_{\beta }^2 s_{\beta }^2+\lambda _4 \chi^2  c_{\beta }^2 s_{\beta }^2)+\rho ^4(\lambda _{ \varphi }  s_{\alpha }^4+\lambda _S c_{\alpha }^4 + \lambda _{S\varphi} c_{\alpha }^2 s_{\alpha }^2)  \\ \nonumber 
&\quad+\rho ^2 r^2(\lambda _{H_1 S} c_{\alpha }^2 c_{\beta }^2 +\lambda _{H_1 \varphi}  c_{\beta }^2 s_{\alpha }^2  +\lambda _{H_2 S}  c_{\alpha }^2 s_{\beta }^2 +2 \lambda _{S \varphi}^\prime \zeta \chi  c_{\alpha } c_{\beta } s_{\alpha } s_{\beta } \cos (\sigma ) + \lambda _{H_2 \varphi}  s_{\alpha }^2 s_{\beta }^2).
\end{align}
where, $\sigma=\Sigma_1+\Sigma_2$ varies between $[0,2\pi]$.
Then, $V_4$ can be written in the matrix form in the basis $R=[r^2,\rho^2]$ as $R^TXR$, with $X_{ij}$ being in components,
\begin{align}
    X_{11}&=\lambda _1 c_{\beta }^4+\lambda _2 s_{\beta }^4 +\lambda _3 c_{\beta }^2 s_{\beta }^2+\lambda _4 \chi ^2 c_{\beta }^2 s_{\beta }^2, \\ 
X_{22}&= \lambda _S c_{\alpha }^4+ s_{\alpha }^4 \lambda _{ \varphi } +\lambda _{S \varphi }c_{\alpha }^2 s_{\alpha }^2, \\ 
X_{12}&=\frac{1}{2} \left( \lambda _{H_1 S} c_{\alpha }^2 c_{\beta }^2 +\lambda _{H_1 \varphi}c_{\beta }^2  s_{\alpha }^2+ \lambda _{H_2 S} c_{\alpha }^2 s_{\beta }^2+\lambda _{H_2 \varphi} s_{\alpha }^2 s_{\beta }^2+2 \lambda _{S \varphi}^\prime \zeta  \chi  c_{\alpha } c_{\beta } s_{\alpha } s_{\beta } \cos (\sigma )  \right).
\end{align}
Thus, $X_{ij}$ is co-positive as far as $X_{ii}\ge0$ and $X_{12}+\sqrt{X_{11}X_{22}}\ge0$.
Depending on the specific values of $\alpha~\rm and~\beta$, the co-positive conditions are simplified to
\begin{align}
&\{ \lambda_1 \ge0,\lambda_S\ge0, \frac{\lambda_{H_1 S}}{2}+\sqrt{\lambda_1 \lambda_S} \ge0\},~\rm for~ \alpha=0~\rm and~\beta=0, \\ 
&\{ \lambda_2 \ge0,\lambda_\varphi \ge0, \frac{\lambda_{H_2 \varphi}}{2}+\sqrt{\lambda_2 \lambda_\varphi } \ge0\},~\rm for~ \alpha=\pi/2~\rm and~\beta=\pi/2, \\ 
&\{ \lambda_1 \ge0,\lambda_\varphi \ge0, \frac{\lambda_{H_1 \varphi}}{2}+\sqrt{\lambda_1 \lambda_\varphi } \ge0\},~\rm for~ \alpha=\pi/2~\rm and~\beta=0, \\
&\{ \lambda_2 \ge0,\lambda_S \ge0, \frac{\lambda_{H_2 S}}{2}+\sqrt{\lambda_2 \lambda_S } \ge0\},~\rm for~ \alpha=0~\rm and~\beta=\pi/2. \\ \nonumber
\end{align}

\subsection{Perturbative unitarity}

We can derive the perturbative unitarity constraints from the calculation of the $2\to 2$ scattering matrix $\mathcal{M}$ by demanding that all the eigenvalues of $\mathcal{M}$ are less than $ 8\pi$.  

We assume that the scattering energy is much higher than the particle masses. Hence, only the contact interactions contribute the $2\to2$ processes. We present the scattering matrices, ${\cal M}_i$, with $i=1,2,\cdots, 5$, for different subsets of scattering processes, in the Appendix B,  and the corresponding set of the eigenvalues, $ \mathcal{E}_i$, for each subset of the scattering processes, read
\begin{align}
 \mathcal{E}_1=&\left\{2 \lambda_1,2 \lambda_2,,2 \lambda_s,2 \lambda_\varphi,\lambda_1+\lambda_2\pm\sqrt{(\lambda_1- \lambda_2)^2+\lambda_4^2}\right\}, \\  
\mathcal{E}_2=&\bigg\{2\lambda_{H_1 S},\lambda_{H_1 \varphi},\lambda_{H_2 S},\lambda_{H_2 \varphi},\frac{1}{2} \left(\lambda_{H_1 \varphi}+\lambda_{H_2 S}\pm\sqrt{(\lambda_{H_1 \varphi} -\lambda_{H_2 S})^2+4 \lambda_{S\varphi}^{\prime 2}}\right), \\
&\quad\frac{1}{2} \left(\lambda_{H_1 S}+\lambda_{H_2 \varphi}\pm\sqrt{(\lambda_{H_1 S} -\lambda_{H_2 \varphi})^2+4 \lambda_{S\varphi}^{\prime 2}}\right)\bigg\}, \\  
\mathcal{E}_3=&\left\{ 2 \lambda_{H_1 S},2 \lambda_{H_2 S} \right\}, \\  
\mathcal{E}_4=&\left\{\lambda_3,\lambda_3+\lambda_4,\lambda_3+\lambda_4,\lambda_3+2 \lambda_4,\pm\sqrt{\lambda_3^2+2 \lambda_3 \lambda_4}\right\}, \\  
\mathcal{E}_5=&\left\{2 \lambda_1,2 \lambda_2,\lambda_3,\lambda_3-\lambda_4,\lambda_3+\lambda_4,\lambda_1+\lambda_2\pm\sqrt{(\lambda_1-\lambda_2)^2+\lambda_4^2} \right\}.
\end{align}

\section{Phenomenological constraints}\label{sec:explimit}

In this section, we discuss the phenomenological constraints coming from neutrino masses, dark matter relic density, DM direct detection experiments, electroweak precision data as well as collider searches.
For dark matter discussion, we assume that the lightest among the DM scalars is a dark matter candidate, which is regarded as being lighter than the RH neutrinos\footnote{The case where the lightest of the RH neutrinos are lighter than the DM scalars was considered in Ref.~\cite{delaVega:2024tuu}}.

\subsection{Neutrino masses}

In the presence of the DM scalar mixings, the relevant interactions for neutrino masses are given by
\bea
{\cal L}_{\rm \nu,{\rm eff}}
&=&- \frac{1}{\sqrt{2}}y_{N,ij} {\bar\nu}_i \Big(\cos\theta_s H_1-\sin\theta_s H_2-i\cos\theta_a A_1+i\sin\theta_a A_2\Big) N_{R,j} \nonumber \\
&&-\frac{1}{2}M_{N,i} \overline{N^c_{R,i}} N_{R,i} +{\rm h.c.} 
\eea
Then, after the $Z_4$ symmetry is broken, we obtain the neutrino masses at one-loop diagrams as
\bea 
({\cal M}_\nu)_{ij}=\frac{1}{16\pi^2}\sum_\alpha\sum_{k} y_{N,ik} y_{N,jk} M_{N,k} F_\alpha \bigg[\frac{m^2_\alpha}{m^2_\alpha-M^2_{N,k}}\ln \frac{m^2_\alpha}{M^2_{N,k}} \bigg]  \label{genumass}
\eea
with $F_\alpha\equiv(\cos^2\theta_s,\sin^2\theta_s,-\cos^2\theta_a, -\sin^2\theta_a )$ and $m_\alpha\equiv(m_{H_1},m_{H_2},m_{A_1},m_{A_2})$.
We note that if the $Z_4$ symmetry is unbroken, $\kappa=\kappa'=0$ and ${\hat m}^2_S=0$, so $\theta_s=\theta_a=0$, for which $H_2, A_2$ do not contribute to neutrino masses and $H_1, A_1$ contributions are cancelled out because of $m_{H_1}=m_{A_1}=m_{H_0}$.

In the limit of decoupling the singlet scalar $S$ with $m_S\gg \kappa', {\hat m}_S$,  the mixing angles for all the dark matter scalars become $|\theta_s|, |\theta_a|\ll 1$ and the mass splittings between dark scalars become small. 
In this case, the neutral inert doublet-like scalars contribute dominantly to the one-loop neutrino masses, so we can approximate eq.~(\ref{genumass})  to
\bea
({\cal M}_\nu)_{ij} \simeq \frac{\lambda_{5,{\rm eff}}v^2_H}{16\pi^2}\sum_k \frac{y_{N,ik}y_{N,jk}M_{N,k}}{m^2_0-M^2_{N,k}}\bigg[1-\frac{M^2_{N,k}}{m^2_0-M^2_{N,k}}\ln\frac{m^2_0}{M^2_{N,k}}\bigg]
\eea
where $m^2_0= m^2_{H_0}$ and  $m^2_{H_1}-m^2_{A_1}\simeq \lambda_{5,{\rm eff}} v^2_H$, with $\lambda_{5,{\rm eff}}$  being given by eq.~(\ref{lambdaeff}). 
Thus, we recover the results in the scotogenic model for neutrino masses where  the singlet scalar $S$ is decoupled and the $Z_2$ symmetry remains \cite{Ma:2006km,Kim:2024cwp}. 

However, for general mixings between dark scalars, the singlet scalar $S$ is not decoupled, so the neutrino masses depend nontrivially on the mixing angles and the scalar spectrum.
In  the limit of the maximal mixings for either CP-even or CP-odd DM scalars, the mass splittings between DM scalars  can be sizable. 

For instance, for the bi-maximal case with $\theta_s=\theta_a=\frac{\pi}{4}$, the one-loop neutrino masses in eq.~({\ref{genumass}}) become
\bea
({\cal M}_\nu)_{ij}&=&\frac{1}{32\pi^2}\sum_{\alpha=H_1,H_2}\sum_{k} y_{N,ik} y_{N,jk} M_{N,k} \bigg[\frac{m^2_{\alpha}}{m^2_{\alpha}-M^2_{N,k}}\ln \frac{m^2_{\alpha}}{M^2_{N,k}} \bigg] \nonumber \\
&&-\frac{1}{32\pi^2}\sum_{\alpha=A_1,A_2}\sum_{k} y_{N,ik} y_{N,jk} M_{N,k} \bigg[\frac{m^2_{\alpha}}{m^2_{\alpha}-M^2_{N,k}}\ln \frac{m^2_{\alpha}}{M^2_{N,k}} \bigg].
\eea
In this case, the DM scalar masses are $m^2_{H_1,H_2}=m^2_{H_0}\pm |\kappa+\kappa'| v_H$ and $m^2_{A_1,A_2}=m^2_{H_0}\pm |\kappa-\kappa'| v_H$.
Then, for $m^2_{H_0}\gg |\kappa\pm \kappa'| v_H$,   we obtain 
\bea
({\cal M}_\nu)_{ij}\simeq \frac{v_H}{16\pi^2}\,(|\kappa+\kappa'|-|\kappa-\kappa'|) \sum_k \frac{y_{N,ik}y_{N,jk}M_{N,k}}{m^2_0-M^2_{N,k}}\bigg[1-\frac{M^2_{N,k}}{m^2_0-M^2_{N,k}}\ln\frac{m^2_0}{M^2_{N,k}}\bigg],
\eea
with  $m^2_0= m^2_{H_0}$. Therefore, we need the different mass splittings for the CP-even and CP-odd scalars for nonzero neutrino masses, as $m^2_{H_1}-m^2_{H_2}=2 |\kappa+\kappa'| v_H\neq m^2_{A_1}-m^2_{A_2}=2 |\kappa-\kappa'| v_H$.

We also consider the case where one of the DM mixing angles is nonzero.
 First, if $\theta_s=\frac{\pi}{4}$ but $|\theta_a|\ll 1$, the neutrino masses become
\bea
({\cal M}_\nu)_{ij}&=&\frac{1}{32\pi^2}\sum_{\alpha=H_1,H_2}\sum_{k} y_{N,ik} y_{N,jk} M_{N,k} \bigg[\frac{m^2_{\alpha}}{m^2_{\alpha}-M^2_{N,k}}\ln \frac{m^2_{\alpha}}{M^2_{N,k}} \bigg] \nonumber \\
&&-\frac{1}{16\pi^2}\sum_{k} y_{N,ik} y_{N,jk} M_{N,k} \bigg[\frac{m^2_{A_1}}{m^2_{A_1}-M^2_{N,k}}\ln \frac{m^2_{A_1}}{M^2_{N,k}} \bigg].
\eea
In this case, the DM scalar masses are $m^2_{H_1,H_2}=m^2_{H_0}\pm |\kappa+\kappa'| v_H$ and $m^2_{A_1}\simeq m^2_{H_0}+(\kappa-\kappa')^2v^2_H/(m^2_s-m^2_a)$. Thus,  for $m^2_{H_0}\gg |\kappa\pm \kappa'| v_H$,  the above results are approximated to
\bea
({\cal M}_\nu)_{ij}=\frac{v^2_H}{8\pi^2}\frac{(\kappa-\kappa')^2}{m^2_a-m^2_s} \sum_k \frac{y_{N,ik}y_{N,jk}M_{N,k}}{m^2_0-M^2_{N,k}}\bigg[1-\frac{M^2_{N,k}}{m^2_0-M^2_{N,k}}\ln\frac{m^2_0}{M^2_{N,k}}\bigg],
\eea
with $m^2_0= m^2_{H_0}$. Thus, we find that neutrino masses are suppressed by the squared mixing angle, $\theta_a^2\simeq \frac{(\kappa-\kappa')^2v^2_H}{(m^2_a-m^2_s)^2}$.
Similarly, if $\theta_a=\frac{\pi}{4}$ but $|\theta_s|\ll 1$, the neutrino masses become
\bea
({\cal M}_\nu)_{ij}&=&\frac{1}{16\pi^2}\sum_{k} y_{N,ik} y_{N,jk} M_{N,k} \bigg[\frac{m^2_{H_1}}{m^2_{H_1}-M^2_{N,k}}\ln \frac{m^2_{H_1}}{M^2_{N,k}} \bigg] \nonumber \\
&&-\frac{1}{32\pi^2}\sum_{\alpha=A_1,A_2}\sum_{k} y_{N,ik} y_{N,jk} M_{N,k} \bigg[\frac{m^2_{\alpha}}{m^2_{\alpha}-M^2_{N,k}}\ln \frac{m^2_{\alpha}}{M^2_{N,k}} \bigg].
\eea
In this case, the DM scalar masses are $m^2_{A_1,A_2}=m^2_{H_0}\pm |\kappa-\kappa'| v_H$ and $m^2_{H_1}\simeq m^2_{H_0}+(\kappa+\kappa')^2v^2_H/(m^2_a-m^2_s)$. Thus, for $m^2_{H_0}\gg |\kappa\pm \kappa'| v_H$,  the neutrino masses are approximated to
\bea
({\cal M}_\nu)_{ij}=\frac{v^2_H}{8\pi^2}\frac{(\kappa+\kappa')^2}{m^2_a-m^2_s} \sum_k \frac{y_{N,ik}y_{N,jk}M_{N,k}}{m^2_0-M^2_{N,k}}\bigg[1-\frac{M^2_{N,k}}{m^2_0-M^2_{N,k}}\ln\frac{m^2_0}{M^2_{N,k}}\bigg],
\eea
with $m^2_0= m^2_{H_0}$.  Similarly, we find that neutrino masses are suppressed by the squared mixing angle, $\theta_s^2\simeq \frac{(\kappa+\kappa')^2v^2_H}{(m^2_a-m^2_s)^2}$. Thus, we find the non-decoupling effects of the singlet scalars manifest in the cases with maximal mixings and the neutrino masses can be suppressed when one of the mixing angles is small.

\subsection{Dark matter relic density}

In the limit of decoupling the singlet scalar $S$, the model resembles the inert doublet dark matter model in which the DM candidate is either $H_1$ or $A_1$.
For a positive  sign of $\lambda_{5,{\rm eff}}$ in eq.~(\ref{eq:massdeiff}), $A_1$ can be lighter than $H_1$, so we can regard it as a dark matter candidate.
A negative sign of  $\lambda_{5,{\rm eff}}$ can be taken instead such that $H_1$ is a dark matter candidate, but there is no phenomenological difference. 

When the singlet scalar $S$ is not decoupled, we need to consider the general mixings between the DM scalars and the dark matter candidate is model-dependent.
In general, the relic density is determined by the mixing quartic couplings and the gauge couplings, and the DM annihilations are possible in the following ways:
\begin{itemize}
\item Self-annihilations: ${\rm DM}\,{\rm DM} \to f{\bar f}, WW, ZZ, h_{1,2} h_{1,2}, h_1 h_2$ \\ 
In the resonance region with the SM Higgs, $M_{\rm DM} \simeq m_{h_1}/2$ and a small Higgs-portal coupling to dark matter is required.
For  $M_{\rm DM} < 80$ GeV, the annihilations into three-body states from off-shell $W$ or $Z$ still occur with large branching fractions.
For $M_{\rm. DM}>100\,{\rm GeV}$, the bosonic channels would be dominant.
\item Co-annihilations: ${\rm DM}\, {\rm DM}'\to f\bar{f}, WW, ZZ, h_{1,2} h_{1,2}, h_1 h_2$ \\
A sufficient contribution from the co-annihilations with heavier DM scalars is possible for $(M_{{\rm DM}'}- M_{\rm DM})/M_{\rm DM} < 0.1$, which requires that a relatively small $\lambda_{5,{\rm eff}}$ in the limit of the inert doublet dark matter. 
\end{itemize}

In all the benchmark models that we consider, the DM annihilation into a pair of singlet-like scalars, $h_2h_2$, is present, due to the extra couplings, $\lambda_{S\varphi}, \lambda_{H_2\varphi}$, that are absent in the inert doublet model.

\subsection{DM direct detection bounds}

Dark matter can scatter with nucleons by both elastic and inelastic processes. 
In the case of the inert double-like dark matter, the inelastic scattering processes are initiated by the exchange of the $Z$ boson, so it can be kinematically forbidden when the mass difference between dark matter and the excited states is greater than around $100$ keV.
If dark matter is singlet-like, the inelastic scattering processes do not contribute to direct detection, so we only have to consider the elastic scattering processes with Higgs exchanges as below.

The coherent elastic scattering between dark matter $\chi$ and nucleus $N$ is mediated by the CP-even Higgs bosons, $h_1$ and $h_2$. 
The corresponding scattering cross section is given by
\begin{align}
\sigma\left(\chi N \to \chi N \right) &= \frac{\mu^2_N (Zf_p+(A-Z)f_n)^2 }{4 \pi m^2_\chi A^2 } \left( \frac{\lambda_{\chi\chi h_1} c_\alpha }{m^2_{h_1}} -\frac{\lambda_{\chi\chi h_2} s_\alpha    }{m^2_{h_2}} \right)^2
\end{align}
where $Z,A-Z$ are the numbers of protons and neutrons in the detector nucleus, $\mu_N$ is the reduced mass for the DM-nucleus system, $f_{p,n}$ are the nucleon form factors, and $\lambda_{\chi\chi h_1},\lambda_{\chi\chi h_2}$, are the trilinear couplings between dark matter and Higgs-like scalars. 

\subsection{Electroweak precision data}

The inert doublet, $H_2$, can contribute the self energy of the SM weak-bosons, so the mass splitting and the mass mixings with singlet $Z_2$-odd scalars  can be  constrained by the electroweak precision measurements.  The  the global electroweak fit for the $S,T$ parameters \cite{ParticleDataGroup:2024cfk} shows
$S_{\rm exp} = -0.05 \pm 0.07$ and
$T_{\rm exp} = 0.00 \pm 0.06$,
for $U = 0$, and the correlation coefficient is given by $\rho_{ST} =0.93$.  
Following Ref.~\cite{Grimus:2008nb}, we obtain the oblique parameters $S$ and $T$ in our model,
\begin{equation}
\begin{split}
 \Delta S &= \frac{1}{24 \pi} \left[ 
 (c_{\theta_s} c_{\theta_a})^2 G\left(m_{H_1}^2,m_{A_1}^2,M_Z^2\right)+(c_{\theta_a} s_{\theta_s})^2 G\left(m_{H_2}^2,m_{A_1}^2,M_Z^2\right) \right. \\ & \left. + (c_{\theta_s} s_{\theta_a})^2 G\left(m_{H_1}^2,m_{A_2}^2,M_Z^2\right) +(s_{\theta_s} s_{\theta_a})^2 G\left(m_{H_2}^2,m_{A_2}^2,M_Z^2\right) \right. \\ & \left.
+c_\alpha^2  \hat{G}\left(M_{h_1}^2,M_Z^2\right) +s_\alpha^2 \hat{G} \left(M_{h_2},M_Z^2\right)+c_{\theta_a}^2 \ln \left(m_{A_1}^2\right) \right. \\ & \left.
 +s_{\theta_a}^2 \ln \left(m_{A_2}^2\right)+s_{\theta_s}^2 \ln \left(m_{H_2}^2\right) +c_{\theta_s}^2 \ln \left(m_{H_1}^2\right) +s_\alpha^2 \ln \left(M_{h_2}^2\right) +c_\alpha^2 \ln \left(M_{h_1}^2\right)\right. \\ & \left.
 -2 \ln \left(m_{H^\pm}^2\right)+\left(2 s_w^2-1\right)^2 G\left(m_{H^\pm}^2,m_{H^\pm}^2,M_Z^2\right)
 -\ln \left(m_h^2\right)-\hat{G} \left(m_h^2,M_Z^2\right)
\right],
\end{split}
\end{equation}
\begin{equation}
\begin{split}
 \Delta T= & \frac{1}{16 \pi M_W^2 s_w^2} \left[ -(c_{\theta_s} c_{\theta_a})^2 F\left(m_{H_1}^2,m_{A_1}^2\right)+c_{\theta_a}^2 F\left(m_{H^\pm}^2,m_{A_1}^2\right)-(c_{\theta_a} s_{\theta_s})^2  
F\left(m_{H_2}^2,m_{A_1}^2\right) \right. \\
  &\left.
-(c_{\theta_s} s_{\theta_a})^2 F\left(m_{H_1}^2,m_{A_2}^2\right) 
-(s_{\theta_s} s_{\theta_a})^2 F\left(m_{H_2}^2,m_{A_2}^2\right)+s_{\theta_a}^2 F\left(m_{H^\pm}^2,m_{A_2}^2\right) \right. \\
  &\left.
+3 s_\alpha^2 \left(F\left(M_Z^2,m_{h_2}^2\right)-F\left(M_W^2,m_{h_2}^2\right)\right)+s_{\theta_s}^2 F\left(m_{H^\pm}^2,m_{H_2}^2\right) +c_{\theta_s}^2 F\left(m_{H^\pm}^2,m_{H_1}^2\right) \right. \\
  &\left.
+3 c_\alpha^2 \left(F\left(M_Z^2,m_{h_1}^2\right)-F\left(M_W^2,m_{h_1}^2\right)\right)-3 \left(F\left(M_Z^2,m_h^2\right)-F\left(M_W^2,m_h^2\right)\right) \right]
\end{split}
\end{equation}
where the loop functions are given in the Appendix C. We perform a $\chi^2$ analysis to check the experimental limit as follows,
\begin{equation}
\chi^2 =  \frac{1}{\left(1-\rho_{ST}^2\right)}\left[\frac{\left(S-S_{\rm exp}\right)^2}{\left( \Delta S_{\rm exp} \right)^2}+\frac{\left(T-T_{\rm exp}\right)^2}{\left( \Delta T_{\rm exp} \right)^2}-\frac{2 \rho_{ST} \left(S-S_{\rm exp}\right)\left(T-T_{\rm exp}\right)}{\Delta S_{\rm exp} \Delta T_{\rm exp}}\right]
\end{equation}
where $S_{\rm exp}$ (or $T_{\rm exp}$) and  $\Delta S_{\rm exp}$ (or $\Delta T_{\rm exp}$) correspond to the experimental mean values and $1\sigma$ uncertainties, respectively, and $S=S_{\rm SM}+\Delta S$, $T=T_{\rm SM}+\Delta T$, with $S_{\rm SM}=T_{\rm SM}=0$.

\subsection{Collider constraints}

\begin{itemize}
    \item LEP: There is no direct search for the inert doublet scalars at the LEP, however the processes such as $e^+ e^- \to H^+ H^-/H_i A_i$ lead to very similar signals as in supersymmetric models. The re-interpretation of LEP search for charginos set the upper limit as $M_{H^\pm}>70$ GeV~\cite{Pierce:2007ut}. Similarly,  the LEP search for neutralinos exclude thw parameter space with $M_{H_1}\le80$ GeV and $M_{A_1}\le100$ GeV provided $|M_{H_1}-M_{A_1}|\ge 8$ GeV~\cite{Lundstrom:2008ai}.
    \item $h\to \gamma\gamma$ decay rate:
    In our model the charged Higgs contributes to the $h\to \gamma\gamma$ rate, which can change the measured value at the LHC.
The signal strength of the SM Higgs for channel $h\to \gamma\gamma$ is given by,
\begin{equation}
\mu_{\gamma\gamma}=\frac{\sigma}{\sigma^{\rm SM}}\frac{{\rm Br}(h\to \gamma\gamma)}{{\rm Br}^{\rm SM}(h\to \gamma\gamma)} =\cos^2\alpha~\frac{{\rm Br}(h\to \gamma\gamma)}{{\rm Br}^{\rm SM}(h\to \gamma\gamma)}.
\end{equation}
We perform a $\chi^{2}$ analysis to check the parameter space consistent with the experimental value $\mu_{\gamma\gamma} = 1.1 \pm 0.06$~\cite{ParticleDataGroup:2024cfk}. The $\chi^{2}$ is  computed as,
\begin{equation}
\chi^{2} =\frac{ \mu^2 - \hat\mu^{2} }{\sigma^+ \sigma^- + (\sigma^+ -\sigma^-)(\mu -\hat\mu)}.
\label{chisq}
\end{equation}    

\item $h\to \rm{invisible}$ decay rate:
The SM Higgs can have invisible decay via $h\to H_i H_i/A_i A_i$ for $M_{H_i,A_i}<M_h/2$.
The upper limit on the invisible branching ratio of Higgs at $95\%$ CL is $\rm Br(h\to \rm{invisible})<0.107$~\cite{ParticleDataGroup:2024cfk}.
\end{itemize}

\section{Benchmark models}

In this section, we consider some benchmark models depending on the mixing angles between the DM scalars and the corresponding neutrino masses in our model. We generate the model files from SARAH-v4.15.2~\cite{Staub:2013tta} and use SPheno-v4.0.3~\cite{Porod:2003um} for calculation of particle spectra and decays, and micrOmegas-v6.0.5~\cite{Alguero:2023zol} for the calculation of dark matter observables.

\begin{table}[t]
\centering
{\small\begin{tabular}{|c|c|c|c|c|c|}
\hline
{Parameters}& Scenario I&Scenario II&{\quad Bi-Maximal } &  {\quad Maximal }&Scenario IV\\
\hline
 $(m_{H_0}/\rm{GeV})^2$  & $31623^2$ & $[1,10^{8}]$&$[1,10^{8}]$  &  [$1 ,10^{8}$]  & [$1 ,10^{8}$]\\      
 $(m_{S}/\rm{GeV})^2$  & $[1,10^8]$ &[$10^{10} ,10^{20}$]&  $m_{2}^2$ &$m_{2}^2$ & [$1 ,10^{8}$]\\   
$\kappa$/GeV & 0 & $-m_S$&$-[10^{-4},1]$ &   $-[10^{-4},1]$  & [$1,10^{6}$]   \\
$\mu$/GeV & $0(-1)$ & 0 & 0&  [$1,10^{4}$] &$10^2$        \\
$v_\varphi$/GeV & 106 & $[10^2,10^9]$& $10^{3}$  & $10^3$  & $10^4$\\
$m_{h_2}$/GeV  &150 & 150& 150        &   150   & 150      \\
\hline
$\lambda_{2}$  & 0 & $[10^{-4},4\pi]$ & $[10^{-4},4\pi]$   & $[10^{-4},4\pi]$ &$[10^{-4},4\pi]$\\
$\lambda_{S}$  & 0  & $[10^{-4},4\pi]$& $[10^{-4},4\pi]$   & $[10^{-4},4\pi]$&$[10^{-4},4\pi]$\\
$\lambda_{H_1\varphi}$  & 0 & 0 & 0   & 0&0      \\
$\lambda_{H_2 S}$  & 0  &$[10^{-4},4\pi]$ & $[10^{-4},4\pi]$   & $[10^{-4},4\pi]$&$[10^{-4},4\pi]$\\
$\lambda_{H_1 S}$  & $[10^{-4},1]$  &$[10^{-4},4\pi]$ & $[10^{-4},4\pi]$   & $[10^{-4},4\pi]$&$[10^{-4},4\pi]$\\
$\lambda_4$  & $-0.01$  & $-[10^{-4},4\pi]$& $-[10^{-6},4\pi]$ & $-[10^{-6},4\pi]$  &$-[10^{-4},4\pi]$ \\
$\lambda_3$ & 0.1& $[10^{-4},4\pi]$&$\lambda_{H_1 S}-\lambda_4$ &  $\lambda_{H_1 S}-\lambda_4$ &$[10^{-4},4\pi]$\\
$\lambda_{S\varphi}^\prime$ & 0 & $[10^{-4},4\pi]$  & \quad $[10^{-7},10^{-3}]$   &  \quad $[10^{-7},10^{-3}]$   & $[10^{-4},4\pi]$ \\
$\kappa^\prime/\rm{GeV}$   & 0& $[0.005,6.2\times10^9]$ &   $[0.5\times 10^{-4},0.5]$  &  $[0.5\times 10^{-4},0.5]$ &$[0.5, 6.2 \times 10^4]$ \\
$\lambda_{S\varphi}$ &$[10^{-2},4\pi]$&$[10^{-4},4\pi]$ &\quad   [$10^{-4} , 1$] & \quad   [$10^{-4} , 1$]  &[$10^{-4} , 1$] \\
$\lambda_{H_2\varphi}$ & $4\pi$ & $[10^{-4},4\pi]$ & $\lambda_{S\varphi}$ & $\lambda_{S\varphi}-4\mu/v_\varphi$ &0 \\
\hline
\end{tabular}
}
\caption{The input parameters in the numerical scan of the benchmark models. The parameter choice  for the maximal scenario is the same as the one for the bi-maximal case, except for $\mu\ne0$.  }
\label{tab:scanranges}
\end{table}

We assume that the mixing angle between the SM Higgs and the dark scalar $\varphi$ is zero, namely, $\sin\alpha=0$, in the following discussion. But, we vary the mixings between dark matter scalars (namely, $H_0-s$ among CP-even scalars or $A_0-a$ between CP-odd scalars), which are determined by  the parameters, $\kappa$ and $\kappa^\prime$. We list the parameter choices for the benchmark models in Table \ref{tab:scanranges}.

In the following discussion, we  consider the case that the RH neutrinos are heavier than the lightest neutral scalar field in our model \footnote{See Ref.~\cite{delaVega:2024tuu} for the case with the RH neutrino dark matter.}.
Then, the candidate for scalar dark matter depends on the mass ordering for the $Z_2$-odd dark sector scalars.
For $m_{H_0}>m_a,m_s$, the lighter singlet-like scalar is  a DM candidate: $A_2$ can be a DM candidate for $\mu>0$, for which $m_a<m_s$;  $H_2$ can be a DM candidate for $\mu<0$, for which $m_s<m_a$;  $H_2$ and $A_2$ are degenerate in mass for $\mu=0$, so the singlet-like complex scalar can be a DM candidate. On the other hand, for $m_{H_0}<m_a,m_s$, either $H_1$ and $A_1$ can be a DM candidate.

\subsection{Scenario I: no dark matter mixings} \label{Scenario-I}

In the case with $\kappa=\kappa^\prime=0$, there is no mixing between dark sector scalars, so neutrino masses vanish, so we need to rely on another mechanism for neutrino masses such as high-scale seesaw mechanism. But, we consider this scenario separately for dark matter physics. 

In this scenario, the mass eigenstates become  $H_1=H_0$, $A_1=A_0$, $ H_2=s$ and $A_2=a$, and the mass eigenvalues are $m_{H_1}=m_{A_1}=m_{H_0}$, $m_{H_2}=m_s$ and $m_{A_2}=m_a$. Taking $\mu<0$ for which $m_{H_0}>m_a, m_s$, we can regard $H_2=s$ as a DM candidate. 

\begin{figure}[t]
\centering
\mbox{
\subfigure[]{\includegraphics[width=0.5\textwidth]{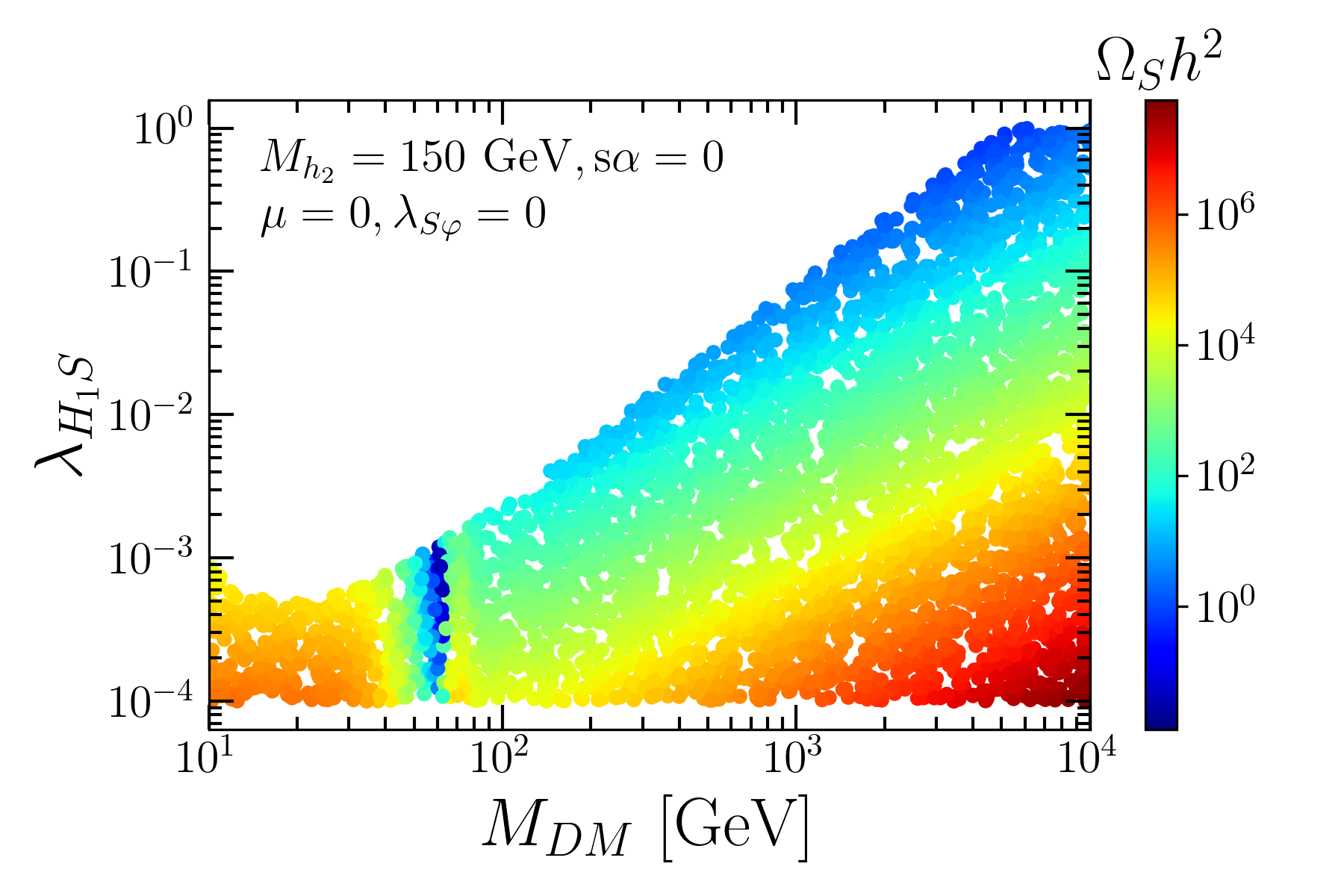}\label{Fig:SSDMa}}
\subfigure[]{\includegraphics[width=0.5\textwidth]{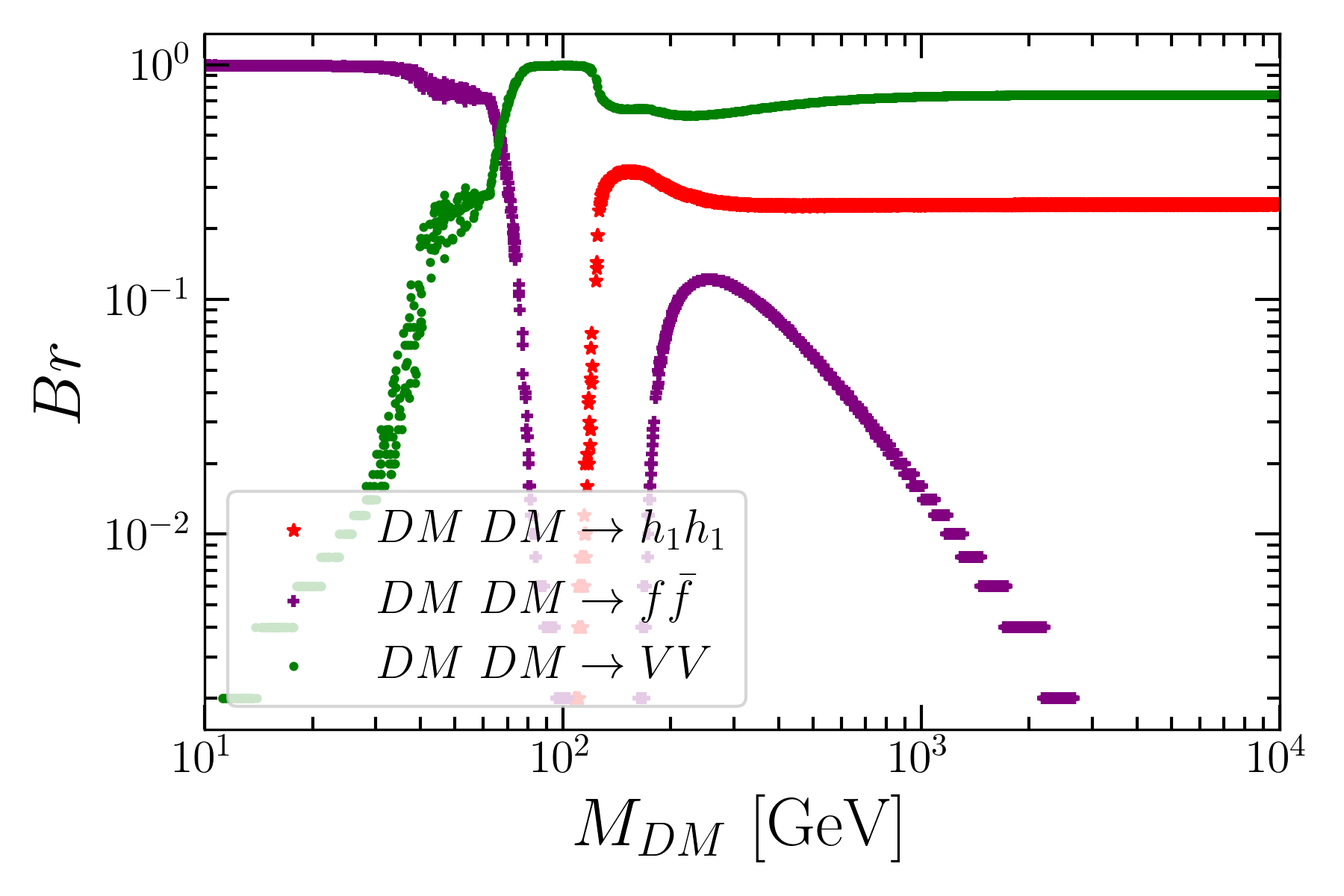}\label{Fig:SSDMb}}
}
\caption{Scenario I: complex singlet dark matter. (a) Variation of the relic density in the plane for $M_{DM}$ vs $\lambda_{H_1 S}$, being consistent with the latest direct detection bound from LZ~\cite{LZ:2024zvo}. (b) Variation of the branching fractions of various channels as a function of DM mass. We chose $\lambda_{S \varphi}=0$ and $\mu=0$.} 
\label{Fig:SSDM}
\end{figure}

If $\lambda_{S\varphi}=0$ and $\mu=0$, the scenario is the same as the complex singlet scalar DM model. In Fig.~\ref{Fig:SSDM}, we show 
 the variation of $\Omega_S h^2$ in the $M_{DM}=M_{H_2}$ vs $\lambda_{H_1 S}$ plane. As evident from the plot, the DM relic density is overabundant in most of the parameter space, because the direct detection bound constrains $|\lambda_{H_2 H_2 h_1}|=v_H |\lambda_{H_1 S}|$ strongly, except in the region with a small $\lambda_{H_1 S}$ near the Higgs resonance. We imposed the current direct detection limit from the LUX-ZEPLIN (LZ) Experiment~\cite{LZ:2024zvo}.  

\begin{figure}[t]
\centering
\mbox{
\subfigure[]{\includegraphics[width=0.5\textwidth]{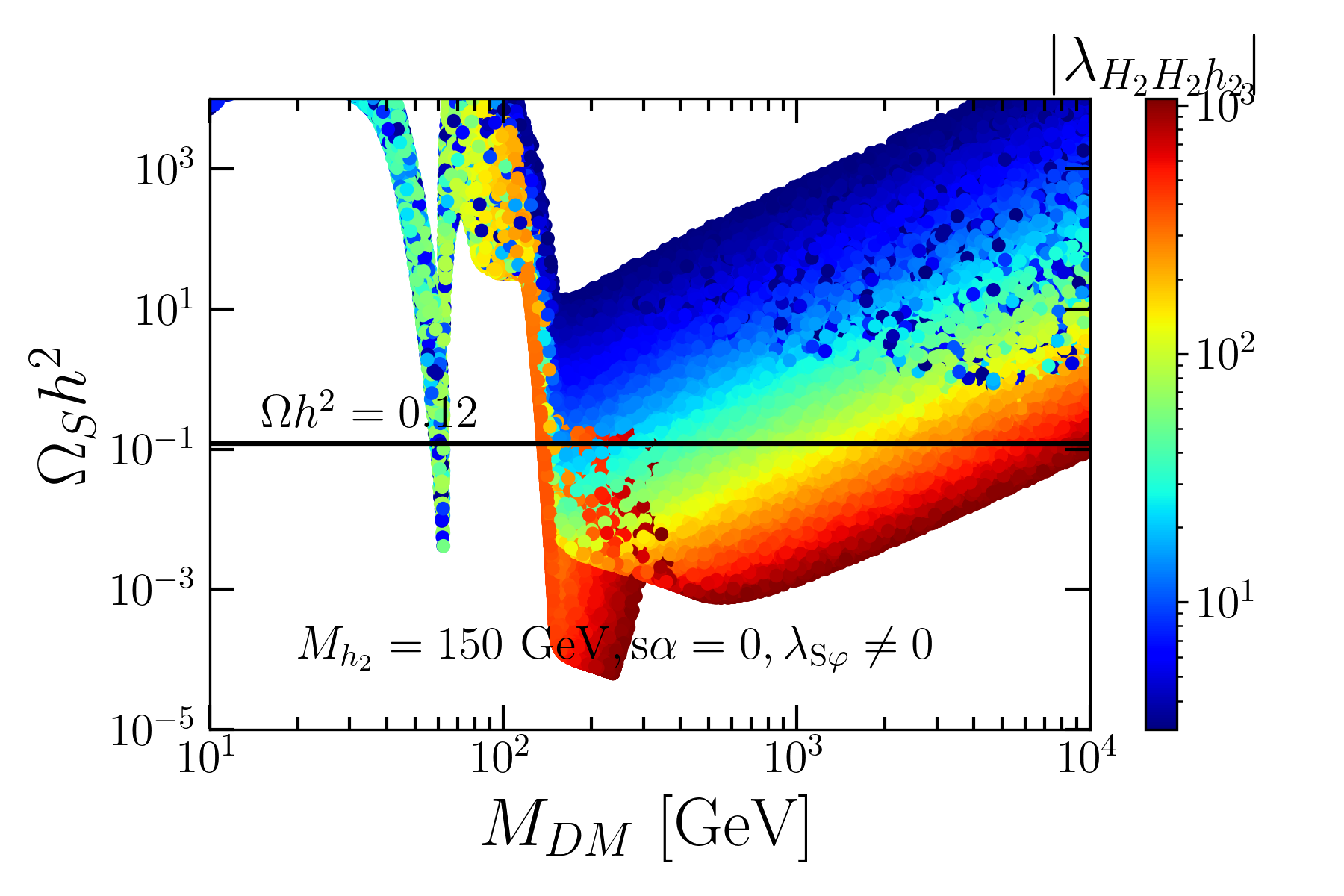}\label{Fig:01a}}
\subfigure[]{\includegraphics[width=0.48\textwidth]{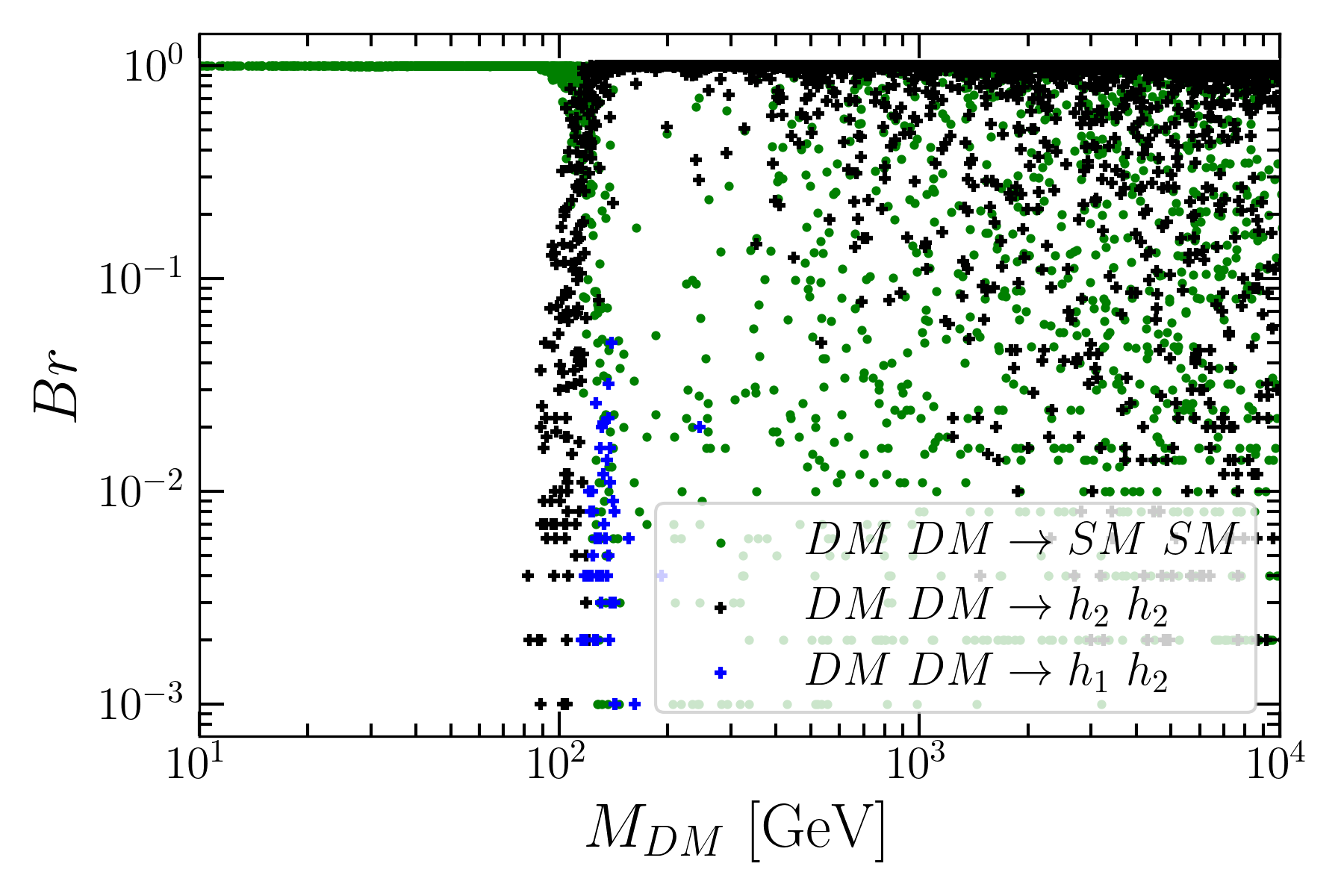}\label{Fig:01b}}
}
\caption{Scenario I: singlet CP-even scalar dark matter. (a)  Variation of the relic density  as a function of $M_{DM}$. The color code indicates the variation of  the $A_2 A_2 h_2$ coupling. (b) Branching fractions of DM annihilation cross sections as a function of $M_{DM}$. We chose a set of the parameters, $\lambda_2=\lambda_S=\lambda_{H_1\varphi}=\lambda_{H_2 S}=0$, $\lambda_3=0.1,\lambda_4=-0.01, \lambda_{H_2 \varphi}=\lambda_\varphi=1$, $\lambda_{H_1 S}=[10^{-4},1]$, and $\lambda_{S \varphi}=[10^{-2},4\pi]$. The mass parameters are taken to $m_2^2=10^9$ GeV, $\mu=-1$ GeV, $m_{h_2}=150$ GeV and $m_{S}^2=[1,10^8]$ GeV.} 
\label{Fig:prel}
\end{figure}

For $\lambda_{S\varphi}\ne 0$ and/or $\mu \ne 0$, if  open kinematically,  the annihilation channel, $H_2 H_2 \to h_2 h_2$ plays an important role to set the observed relic density. To explore this, we consider the parameters as $\mu=-1$ GeV and $m_{h_2}=150$ GeV, $\lambda_{H_1 S}=[10^{-4},1]$ and $\lambda_{S \varphi}=[10^{-2},4\pi]$. The values of other model parameters are given in the second column of Table.~\ref{tab:scanranges}.
We show the relic density as a function of $M_{DM}$ for the varying effective coupling, $|\lambda_{H_2 H_2 h_2}|=|2\mu-\lambda_{S\varphi}v_\varphi|$, in Fig.~\ref{Fig:01a}. We show the observed relic density value by the black line. The chosen parameter space satisfies the DD limit, EW precision data as well as theoretical constraints  such as perturbative unitarity and vacuum stability. The charged Higgs scalar does not alter the $h\to \gamma \gamma$ rate as it is considered to be heavier than $10^4$ GeV. 

For $M_{DM}\ge M_{h_2}$, the channel $H_2 H_2 \to h_2 h_2$ opens up and can lead the DM annihilation process for a sizable coupling $\lambda_{S\varphi}$. This channel is mediated via contact interaction $\lambda_{S\varphi}$ as well as via  trilinear interactions $\lambda_{H_2 H_2 h_2}$ and $\lambda_{h_2 h_2 h_2}$ in the  $s,t,u$ channel processes. Depending on the size of the effective vertex factor, there exists a cancellation among the different diagrams. This is evident in the region of $150~\rm GeV<M_{DM}<400~\rm GeV$ where contributions of various diagrams are comparable. For $M_{DM}>400~\rm GeV$, contributions from $s,t,u$ channels are suppressed by $1/M_{DM}^2$ and the leading contribution is from the contact interaction. Note that this cancellation region shifts depending upon the choice of parameters. As can be seen from this plot, there exists a large parameter space where $\Omega_S h^2\le0.12$ while satisfying all other bounds. 

In Fig.~\ref{Fig:01b}, we present the variation of  the branching fractions of the DM annihilation cross sections as a function of $M_{DM}$. The channel $H_2 H_2 \to h_2 h_2$ can give rise to almost $100\%$ contribution to the DM annihilation cross sections as far as it is open. We note that the next to lightest-odd-particle (NLOP), $A_2$ contributes to the DM annihilation cross sections if the relative mass splitting is small enough, namely, $(M_{DM}-M_{A_2})/M_{DM}\lesssim0.1 $.

\begin{figure}[t]
\centering
\mbox{
\subfigure[]{\includegraphics[width=0.5\textwidth]{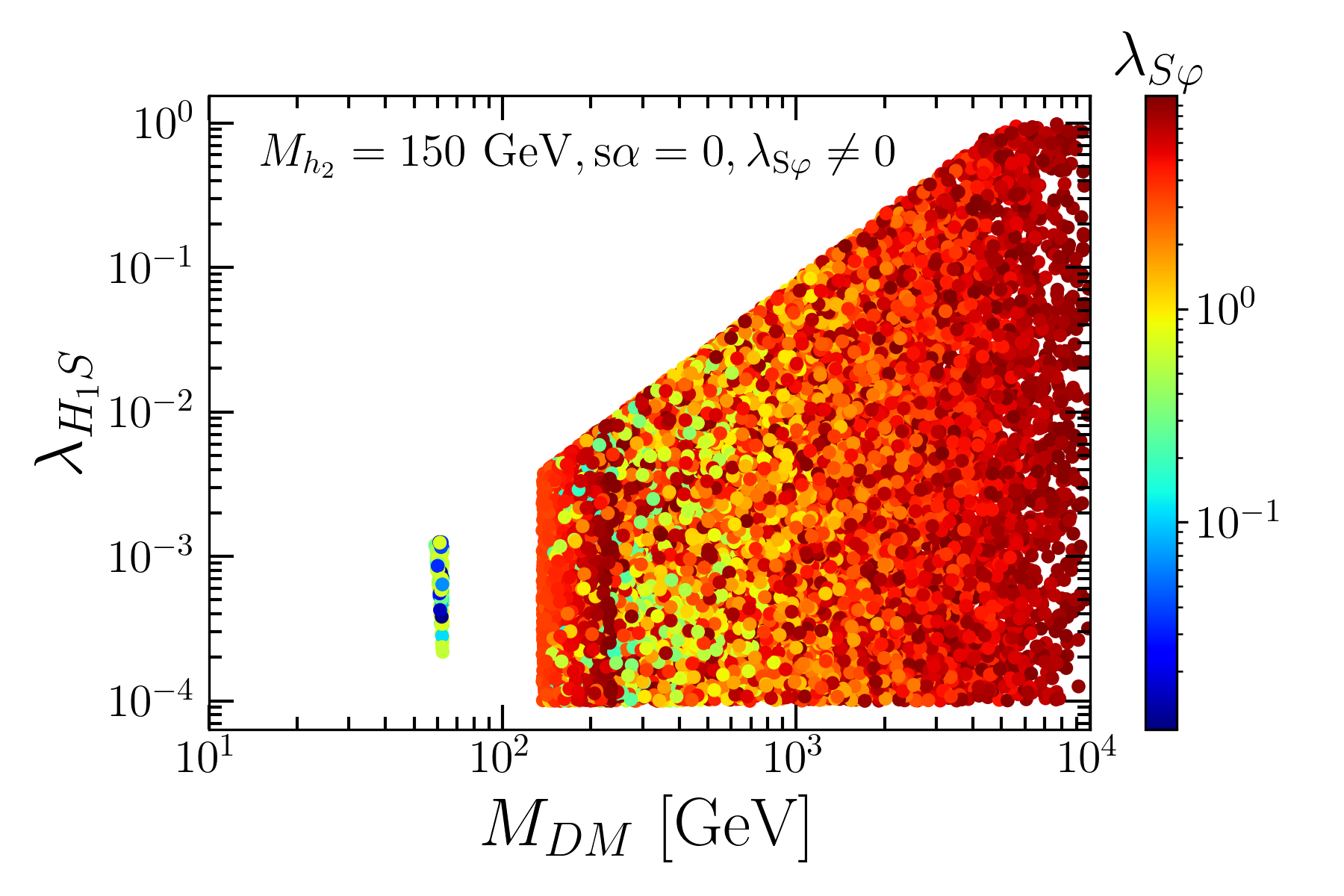}}
}
\caption{Scenario I: singlet CP-even scalar dark matter. Parameter space satisfying $1\%-100\%$ of the observed relic density. The choice of the parameters is the same as in Fig.~\ref{Fig:prel}.} 
\label{Fig:1b}
\end{figure}
    
In Fig.~\ref{Fig:1b}, varying $\lambda_{S \varphi}$ in the range of $\lambda_{S \varphi}=[0.1,4\pi]$, we  present the allowed parameter space for  $M_{DM}=M_{H_2}$ vs $\lambda_{H_1 S}$  where both the observed relic density and direct detection bound can be satisfied simultaneously. If $H_2 H_2 \to h_2 h_2$ is open kinematically, it can drive the DM annihilation for suitable (non-zero) values of $\lambda_{S \varphi}$. For $m_{h_2}=150$ GeV, we obtain  $1\%-100\%$ of the observed relic density in the range of $M_{DM}=[150,10^4]$ GeV for $\lambda_{S \varphi}=[0.1,4\pi]$, being consistent with perturbativity. 
The annihilation channel, $H_2 H_2 \to h_2 h_2$, does not provide a notable contribution  if $\lambda_{S \varphi}<0.1$.

\subsection{Scenario II: small dark matter mixings} \label{Scenario II}
Next, as discussed in Section 3.3, we consider the case where $\kappa, \kappa'$ are nonzero but smaller than $m_S$.
In this case, as discussed in Section 4.1, small neutrino masses are generated, due to a small $\lambda_{5,{\rm eff}}$. 

In this scenario, as $m_s,m_a\gtrsim m_{H_0},m_{A_0}$, the model becomes similar to the inert doublet model with a small mass splitting between $H_1$ and $A_1$ in eq.~(\ref{eq:massdeiff}), which is approximated to 
\bea
  m^2_{H_1}-m^2_{A_1} \simeq  -\frac{4\kappa\kappa' v^2_H}{m^2_S}.
 \eea
 Then, for $\kappa\kappa'<0$, $A_1$ is a dark matter candidate.
 
 If $\lambda_{H_2 \varphi}=0$, the DM phenomenology is the same as in the inert doublet model. But, for $\lambda_{H_2 \varphi}\ne 0$,  the extra annihilation channel, $A_1 A_1 \to h_2 h_2$, plays an important role to set the observed relic density, if  open kinematically. 

\begin{figure}[t]
\centering
\mbox{
\subfigure[]{\includegraphics[width=0.5\textwidth]{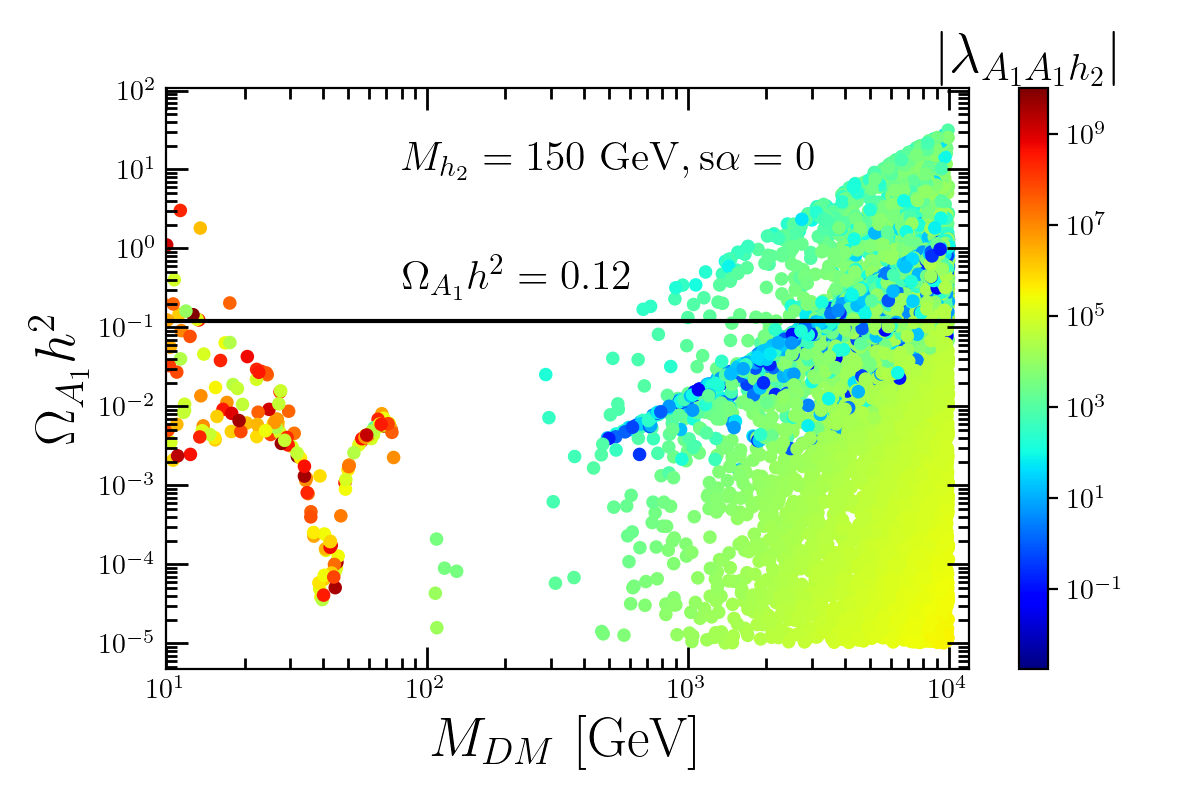}\label{Fig:02a}}
\subfigure[]{\includegraphics[width=0.48\textwidth]{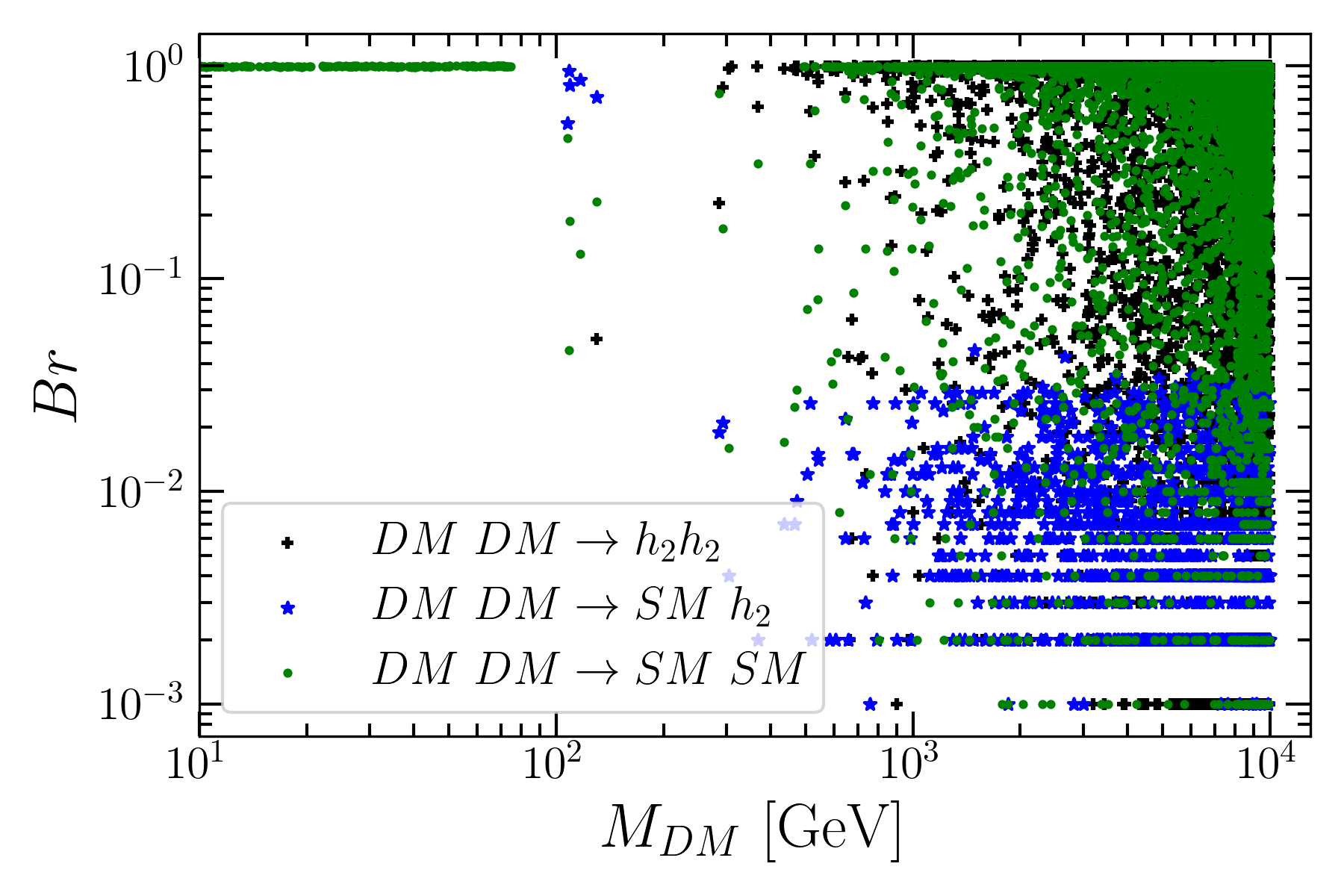}\label{Fig:02b}}
}
\caption{Scenario II: doublet-like CP-odd scalar dark matter. (a) The relic density as a function of $M_{DM}$. The color code indicates the variation of  the $A_1 A_1 h_2$ coupling. (b) Branching fractions of DM annihilation cross sections as a function of $M_{DM}$. We chose a set of the parameters, $\lambda_{H_1\varphi}=0, \lambda_2=\lambda_S=\lambda_{H_1 S}=\lambda_{H_2 S}=\lambda_{S \varphi}^\prime=\lambda_3=-\lambda_4= [10^{-4},4 \pi]$, and $\lambda_{H_2 \varphi}= [10^{-4},4 \pi]$. The mass parameters are taken to $m_S^2=[10^{10},10^{20}]$ GeV, $\mu=0$ GeV,  $\kappa=-m_S$, and $m_{h_2}=150$ GeV and $m_{H_0}^2=[1,10^{8}]$ GeV and $v_{\varphi}=[10^{2},10^{9}]$ GeV. } 
\label{Fig:0doublet}
\end{figure}

 To analyze this scenario, we fix the parameters, as given in the  third column of Table.~\ref{tab:scanranges}.
 As a result, in Fig.~\ref{Fig:02a}, we show the variation of the relic density as a function of the $M_{DM}$ in the scenario II. Here, the color code indicates the variation of  effective DM coupling, given by $|\lambda_{A_1 A_1 h_2}|\simeq |\lambda_{H_2\varphi}| v_\varphi$. In Fig.~\ref{Fig:02b}, we also show the variation of the branching fractions of DM annihilation cross sections as a function of $M_{DM}$ for the same scenario.
 
\begin{figure}[t]
\centering
\mbox{
\subfigure[]{\includegraphics[width=0.5\textwidth]{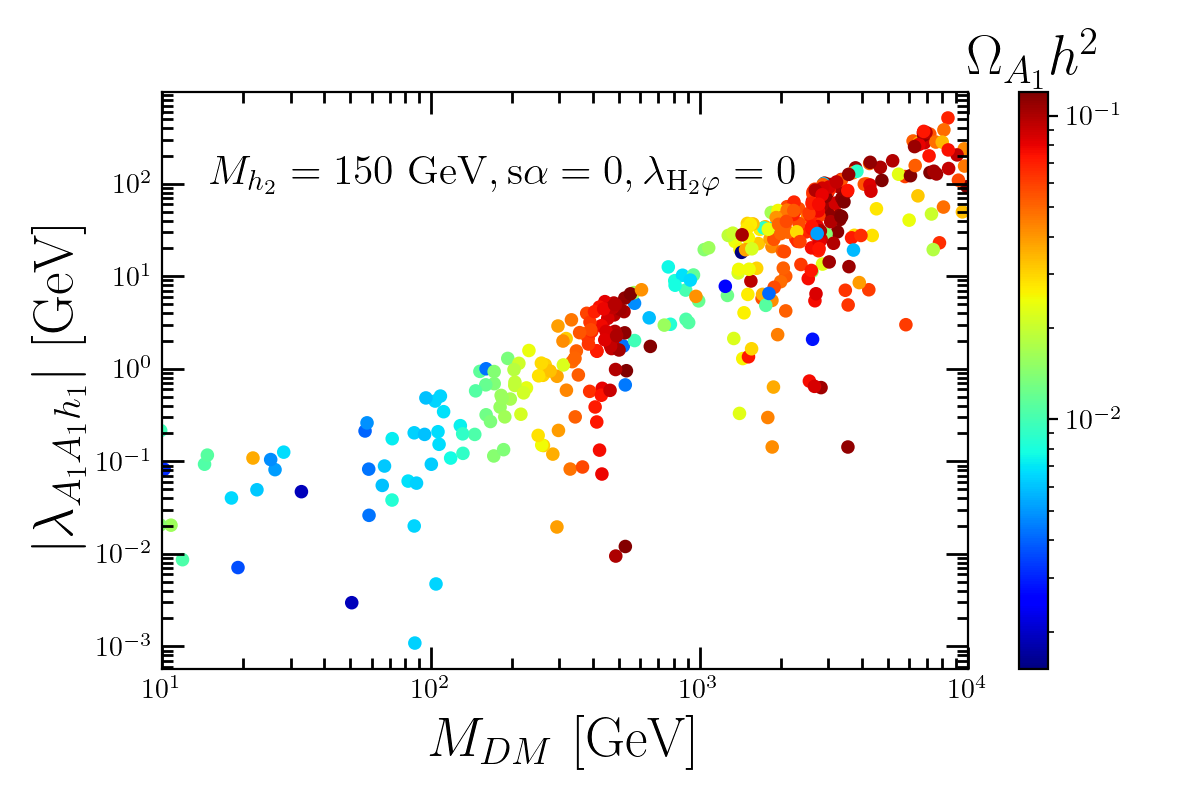}\label{Fig:2a}}
\subfigure[]{\includegraphics[width=0.5\textwidth]{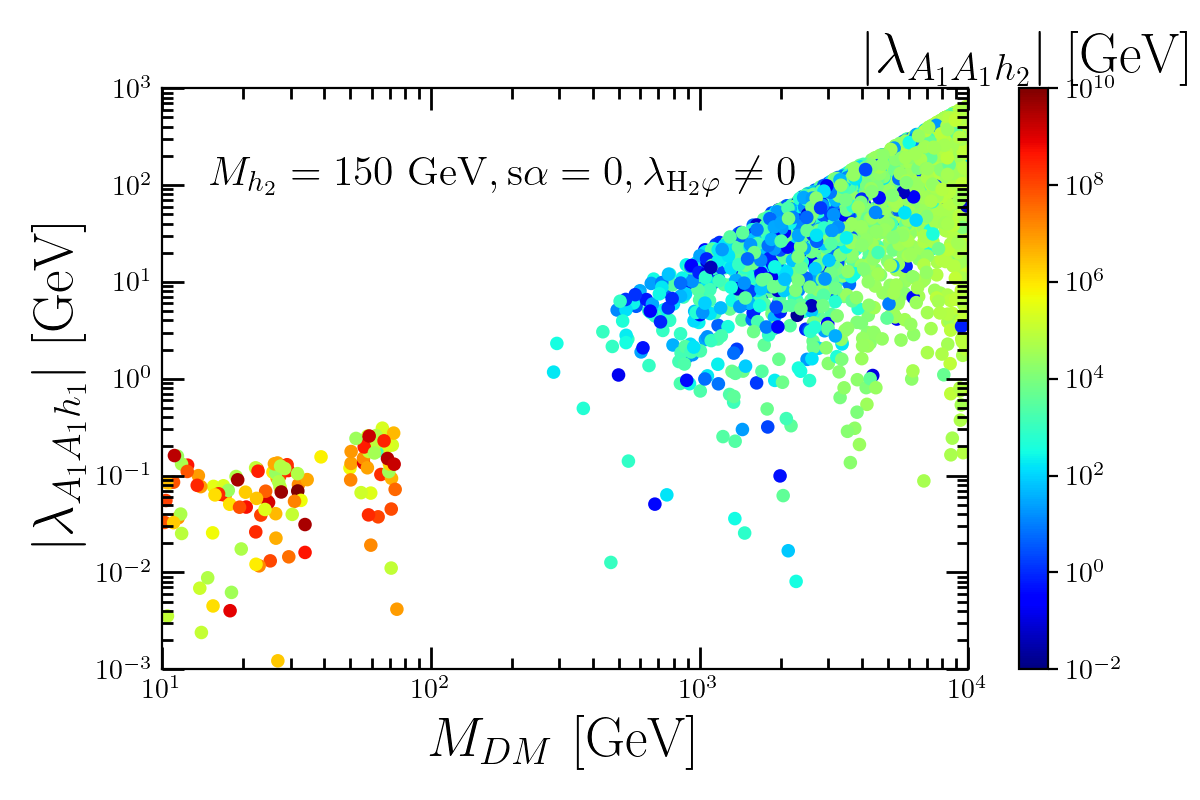}\label{Fig:2b}}
}
\caption{Scenario II: doublet-like CP-odd scalar dark matter. (a) Variation of the relic density in the  plane for $M_{DM}$ vs $|\lambda_{A_1 A_1 h_1}|$, for $\lambda_{H_2 \varphi}=0$. The parameter space is consistent with the direct detection bound and $100 \% ~\rm{to}~1\%$ of the observed DM relic density is explained. (b) The same as in (a), but for  $\lambda_{H_2 \varphi}\neq 0$. The choice for the other parameters  for both (a) and (b) is the same as in Fig.~\ref{Fig:0doublet}.} 
\label{Fig:relicdoublet}
\end{figure}

In Fig.~\ref{Fig:2a}, we present the  variation of $\Omega_{A_1} h^2$ in the $M_{DM}=m_{A_1}$ vs $|\lambda_{A_1 A_1 h_1}|$ plane, assuming 
$\lambda_{H_2 \varphi}=0$. Here,  we demanded $A_1$ to build up $1\%-100\%$ of the observed relic density while being consistent with the direct detection bound on $|\lambda_{A_1 A_1 h_1}|=|\lambda_L| v_H$ with $\lambda_L=\lambda_3+\lambda_4$. As evident from the plot, the DM relic density is overabundant for $M_{\rm DM}\gtrsim 10^3$ GeV and  $|\lambda_{A_1 A_1 h_1}|<1$ GeV.  
The reason is the following. For heavy DM masses $\gtrsim 1\,{\rm TeV}$, the DM annihilations into longitudinal components of W, Z bosons are dominant.
According to the Goldstone equivalence theorem (for $\sqrt{s}\simeq 2 m_{A_1} \gg m_{W,Z}$), $A_1 A_1 \to Z_L Z_L$ is proportional to $\lambda_L$ and $A_1 A_1 \to W_L W_L$ is proportional to $ \lambda_4$ in our case.
Since $\lambda_L$ is bounded by direct detection whereas $\lambda_4$ is bounded by electroweak precision data (as it leads to a mass splitting between $m_{H^\pm}$ and $m_{H_0, A_0}$ within the inert doublet).
So, both $A_1 A_1\to  Z_L Z_L$ and $A_1 A_1\to W_L W_L$ are suppressed for DM mass greater than $1\,{\rm TeV}$.

In Fig.~\ref{Fig:2b}, we present the allowed parameter space in the same parameter space as in Fig.~\ref{Fig:2a}, but for $\lambda_{H_2 \varphi} \ne 0$. If $A_1 A_1 \to h_2 h_2$ is open kinematically, it can drive the DM annihilation for suitable (non-zero) values of $\lambda_{H_2 \varphi}$. For $m_{h_2}=150$ GeV, we obtain $1\%-100\%$ of the observed relic density for $\lambda_{H_2 \varphi}\ne0$ and $M_{DM}>10^3$ GeV, while the relic density is overabundant for $\lambda_{H_2 \varphi}=0$. The annihilation channel, $A_1 A_1 \to h_2 h_2$, opens up again a wider parameter space for $\lambda_{H_2 \varphi} \ne 0$.

\subsection{Scenario III: maximal dark matter mixings}
 
Next, we consider the case where $\theta_s$ and/or $\theta_a$ are equal to $\frac{\pi}{4}$. In this scenario, as discussed in Section 4.1, some (or all) the diagonal masses for DM scalars are degenerate for either $\theta_s$ or $\theta_a$ equal to $\frac{\pi}{4}$ ($\theta_s=\theta_a=\frac{\pi}{4}$), so neutrino masses can be suppressed by small mass splitting(s) between dark matter scalars.

For $\theta_s=\frac{\pi}{4}$ and $|\theta_a|\ll 1$ (or $\theta_a=\frac{\pi}{4}$ and $|\theta_s|\ll 1$), we find that $m_{H_0}=m_s$ (or $m_{H_0}=m_a$). In each case with maximal dark matter mixing(s), a combination of singlet and inert doublet models becomes a dark matter candidate.
A singlet-like scalar $A_2$ ($H_2$) is a DM candidate for $\mu>0$ ($\mu<0$). In the maximal cases, the effective DM couplings are given by
\bea
|\lambda_{A_2A_2 h_2}|&=& \bigg|\frac{1}{2}(2\mu-\lambda_{S\varphi}v_\varphi) -\frac{1}{2}\lambda_{H_2\varphi} v_\varphi -\frac{\kappa'}{v_\varphi}\,v_H\bigg|, \label{lamA2h2} \\
|\lambda_{A_2 A_2 h_1}|&=& \bigg|\frac{1}{2} (\lambda_3+\lambda_4) v_H+\frac{1}{2}\lambda_{H_1S} v_H -\kappa+\kappa'  \bigg|,  \label{lamA2h1} 
\eea
for $\mu>0$, $\theta_a=\frac{\pi}{4}$ and $|\theta_s|\ll 1$;
\bea
|\lambda_{H_2H_2 h_2}|&=& \bigg|\frac{1}{2}(2|\mu|-\lambda_{S\varphi}v_\varphi) -\frac{1}{2}\lambda_{H_2\varphi} v_\varphi +\frac{\kappa'}{v_\varphi}\,v_H \bigg|,   \label{lamH2h2} \\
|\lambda_{H_2 H_2 h_1}|&=& \bigg|\frac{1}{2} (\lambda_3+\lambda_4) v_H+\frac{1}{2}\lambda_{H_1S} v_H -\kappa-\kappa'  \bigg|,  \label{lamH2h1} 
\eea
for $\mu<0$, $\theta_s=\frac{\pi}{4}$ and $|\theta_a|\ll 1$.
On the other hand, for $\theta_s=\theta_a=\frac{\pi}{4}$, we have  $m_{H_0}=m_a=m_s$, so $\mu=0$. Then, for $\kappa\kappa'<0$ ($\kappa\kappa'>0$), $A_2$ ($H_2$) is a DM candidate. In this case, the corresponding effective DM couplings are given by (\ref{lamA2h2}) and (\ref{lamA2h1}) with $\mu=0$ in the former case and (\ref{lamH2h2}) and (\ref{lamH2h1}) with $\mu=0$ in the latter case.
 In Table~\ref{tab:scanranges}, we present the parameter choices for the  bi-maximal and maximal cases in the fourth and fifth columns, respectively.
 
\begin{figure}[t]
\centering
\mbox{
\subfigure[]{\includegraphics[width=0.5\textwidth]{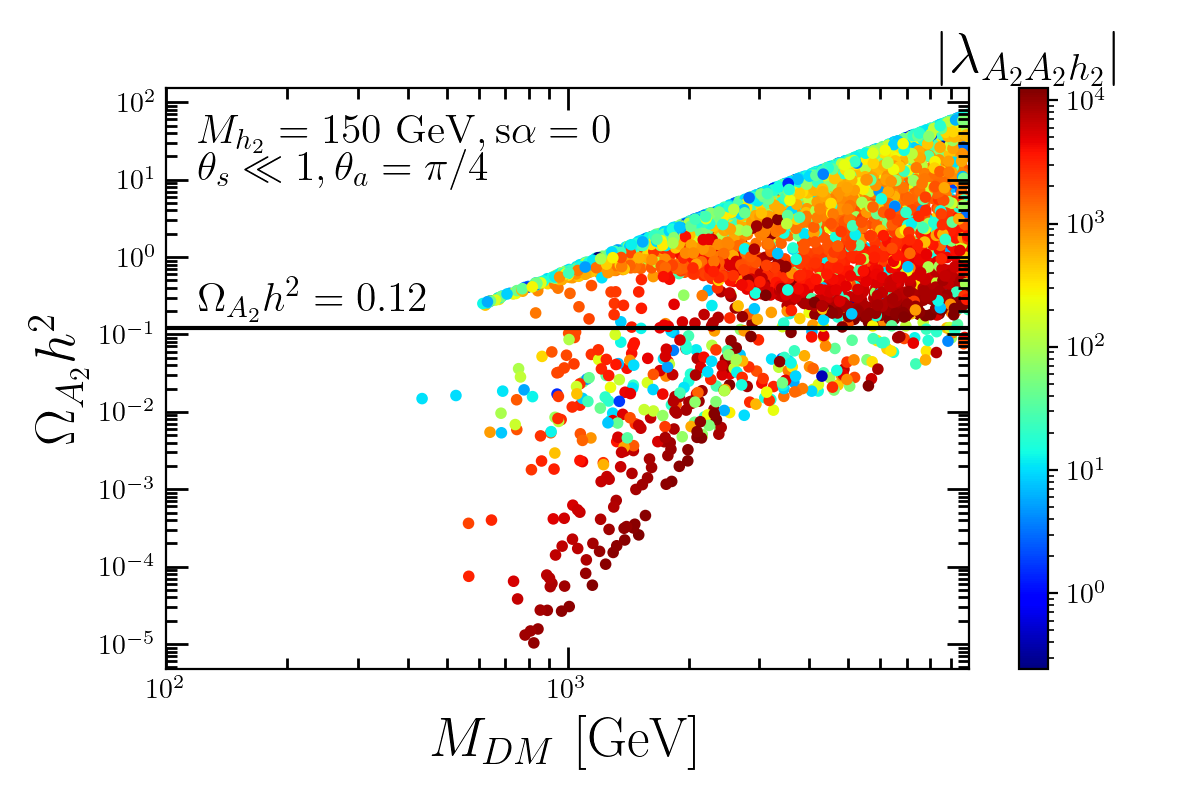}\label{Fig:3a}}
\subfigure[]{\includegraphics[width=0.5\textwidth]{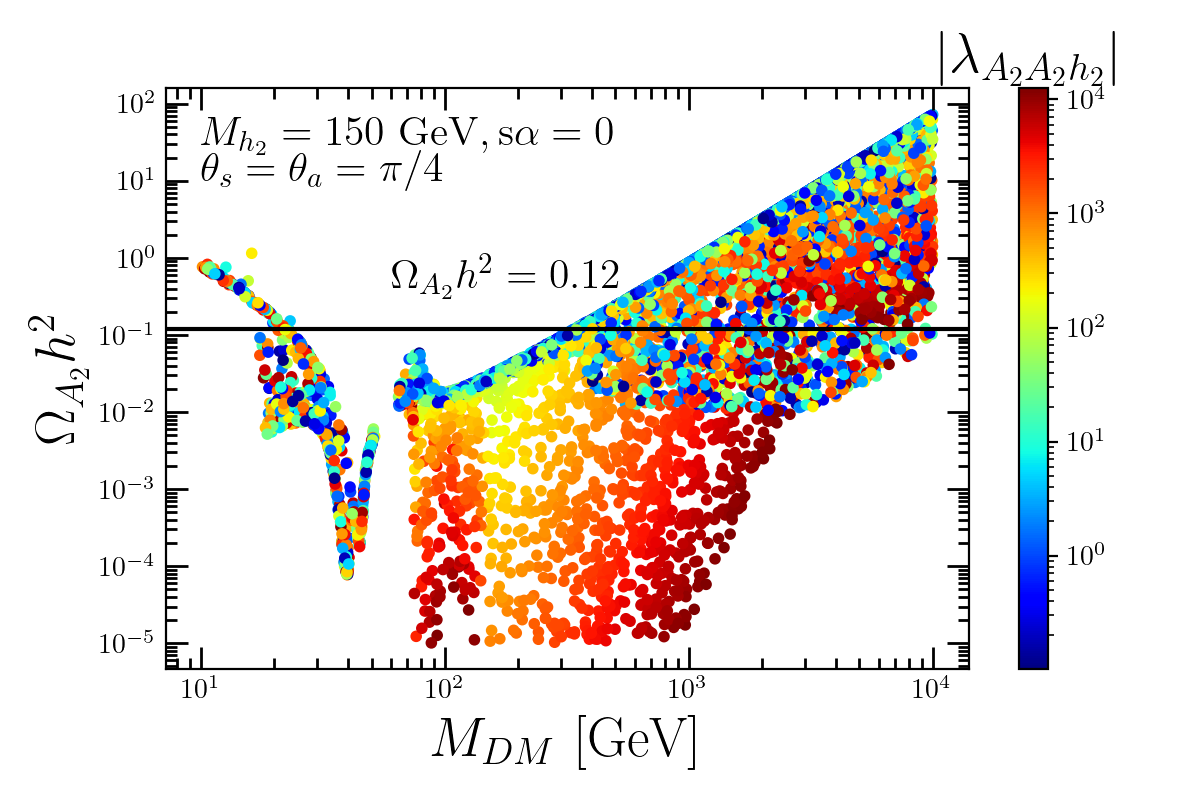}\label{Fig:3b}}
}
\caption{Scenario III: singlet-like CP-odd scalar dark matter. The relic density as a function of  $M_{DM}$: (a) for  the maximal case and (b) for the  bi-maximal case.}
\label{Fig:maximal}
\end{figure}

\begin{figure}[t]
\centering
\mbox{
\subfigure[]{\includegraphics[width=0.48\textwidth]{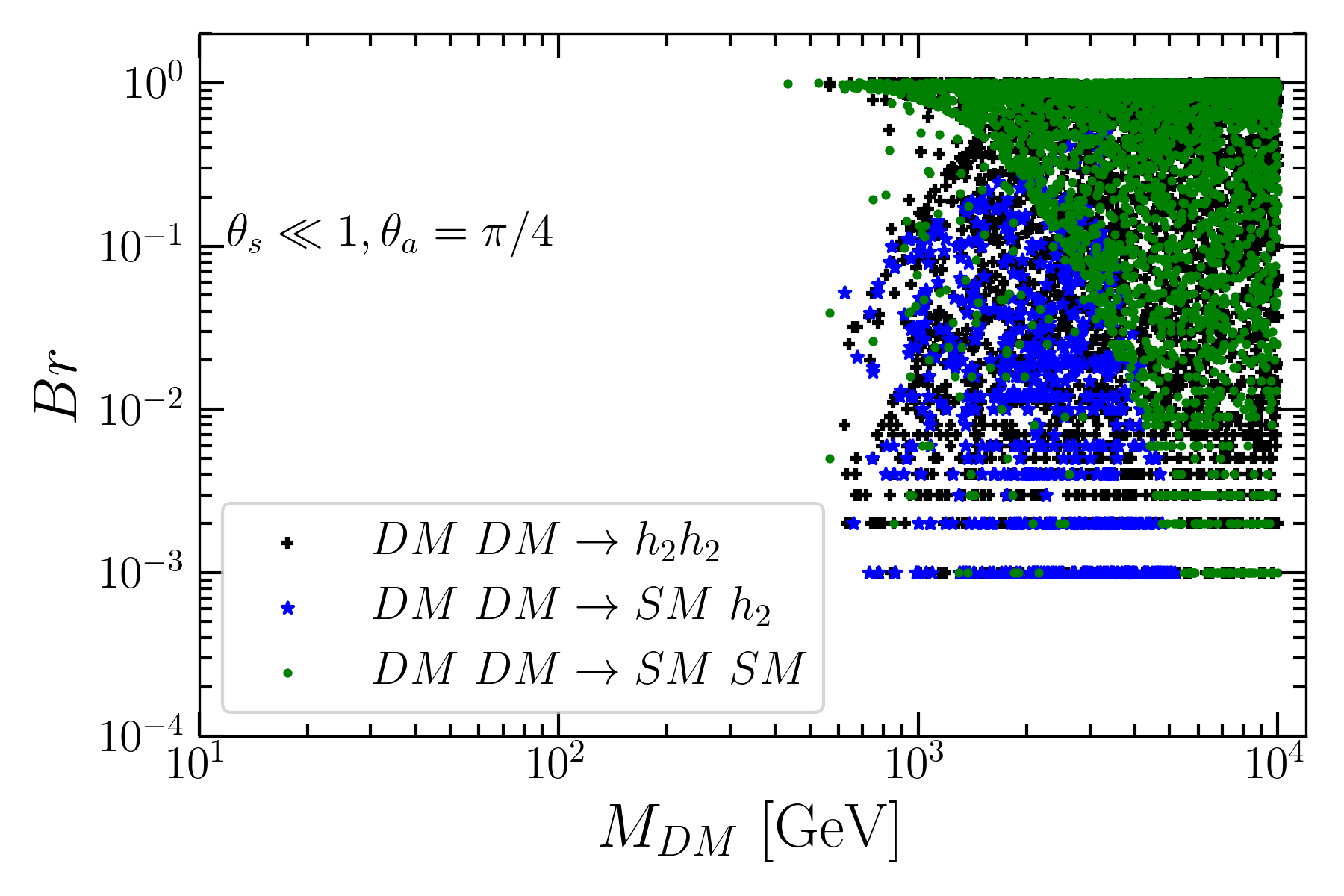}\label{Fig:4a}}
\subfigure[]{\includegraphics[width=0.48\textwidth]{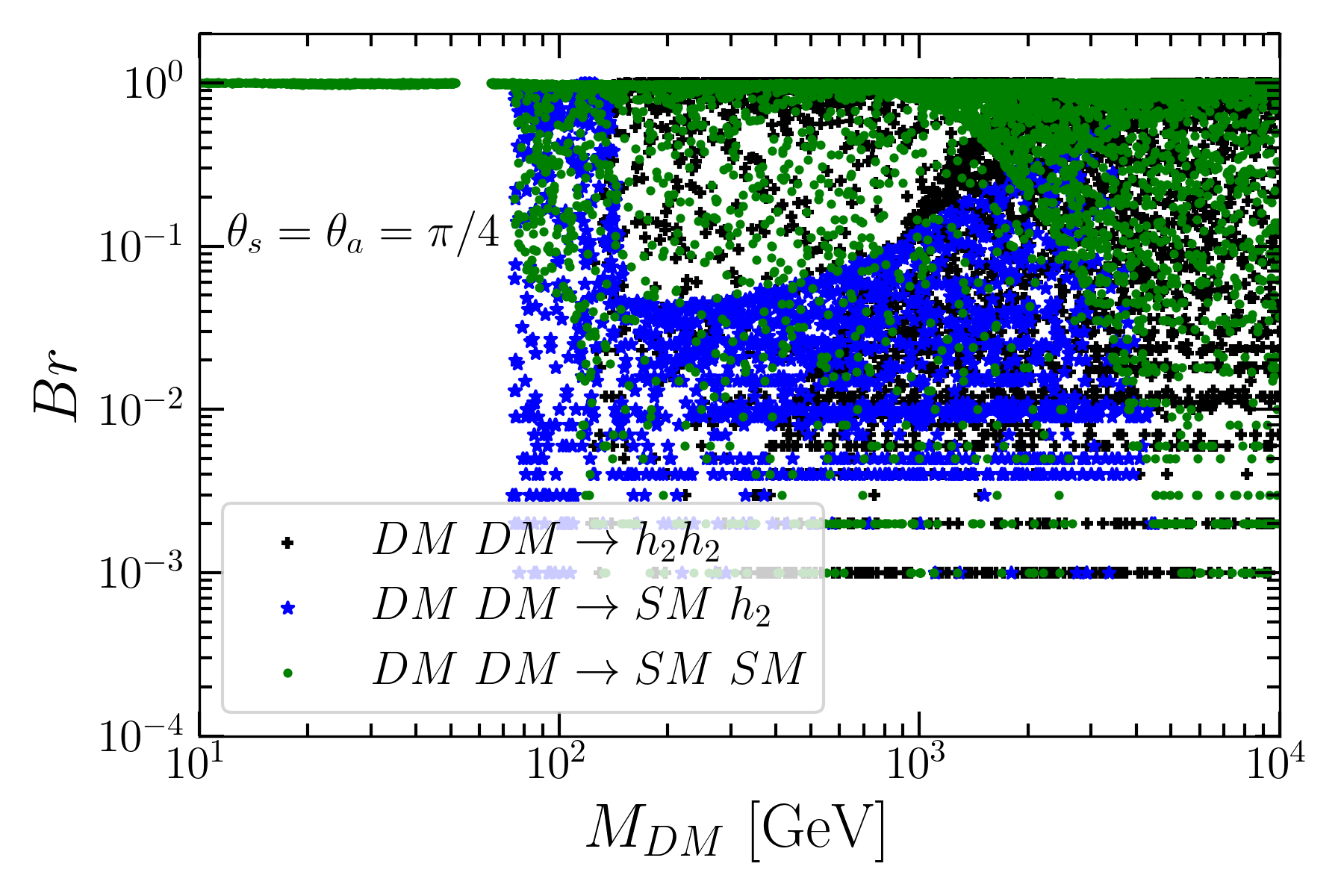}\label{Fig:4b}}
}
\caption{Scenario III: singlet-like CP-odd scalar dark matter. Branching fractions of DM annihilation cross sections as a function of the relative mass splitting between DM and charged Higgs scalars: (a) for  the maximal case and (b) for the bi-maximal case.}
\label{Fig:maximal_2}
\end{figure}

In Fig.~\ref{Fig:3a} and Fig.~\ref{Fig:3b}, we show the variation of the relic density as a function of $M_{DM}$, for the varying effective DM interaction $\lambda_{A_2 A_2 h_2}$, in the maximal and bi-maximal cases, respectively. This parameter space is consistent with all the experimental and theoretical constraints including the direct detection limit. In Fig.~\ref{Fig:4a} and Fig.~\ref{Fig:4b}, we show the variation of the branching fractions of DM annihilation cross sections as a
function of its mass.

\subsection{Scenario IV: large mass gap for singlet dark scalars}

In this scenario, we consider a large value of $\hat{m}_S^2=2\mu v_\varphi$ for which there is a large mass gap between the components of the singlet scalar $S$. For instance, we choose $\mu=10^2$ GeV and $v_\varphi=10^4$ GeV. Then, the lightest scalar among the dark sector scalars is $A_2$ and the heaviest one is $H_2$. The parameter choice for the scenario IV is also given in the last column of Table~\ref{tab:scanranges}. 

\begin{figure}[t]
\centering
\mbox{
\subfigure[]{\includegraphics[width=0.48\textwidth]{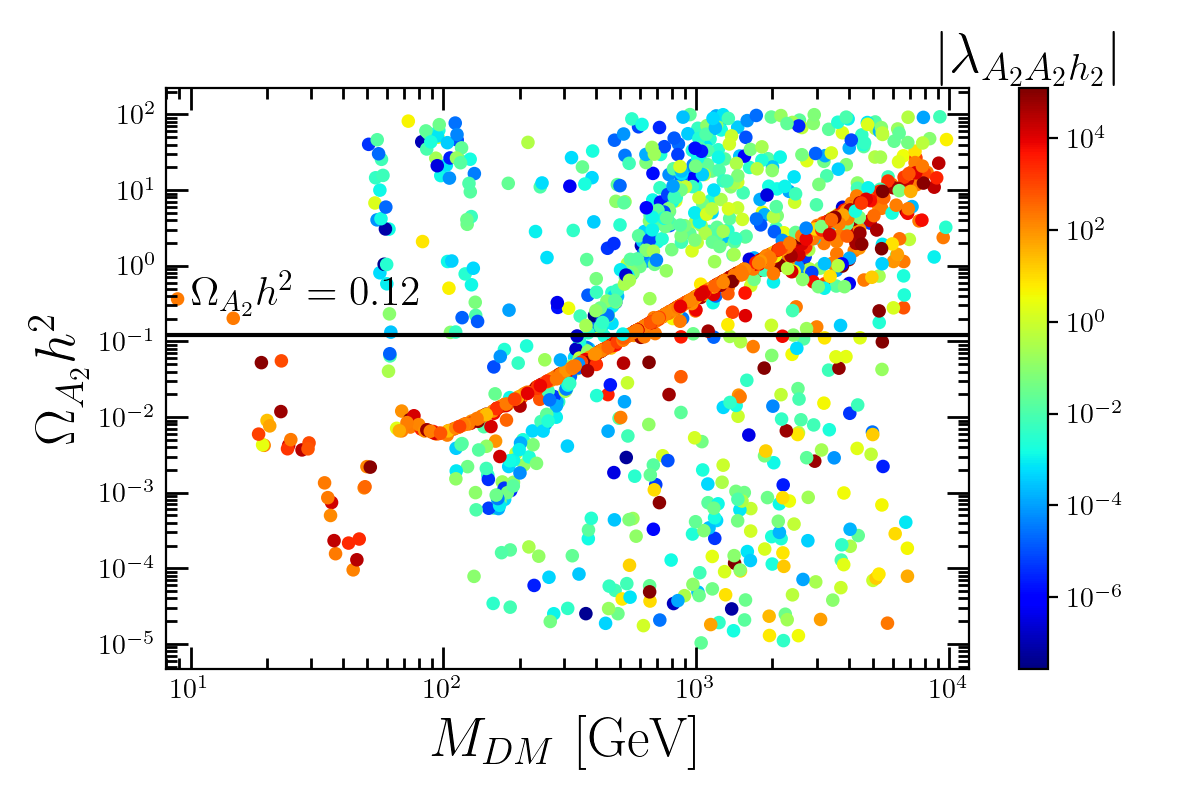}\label{Fig:8a}}
\subfigure[]{\includegraphics[width=0.48\textwidth]{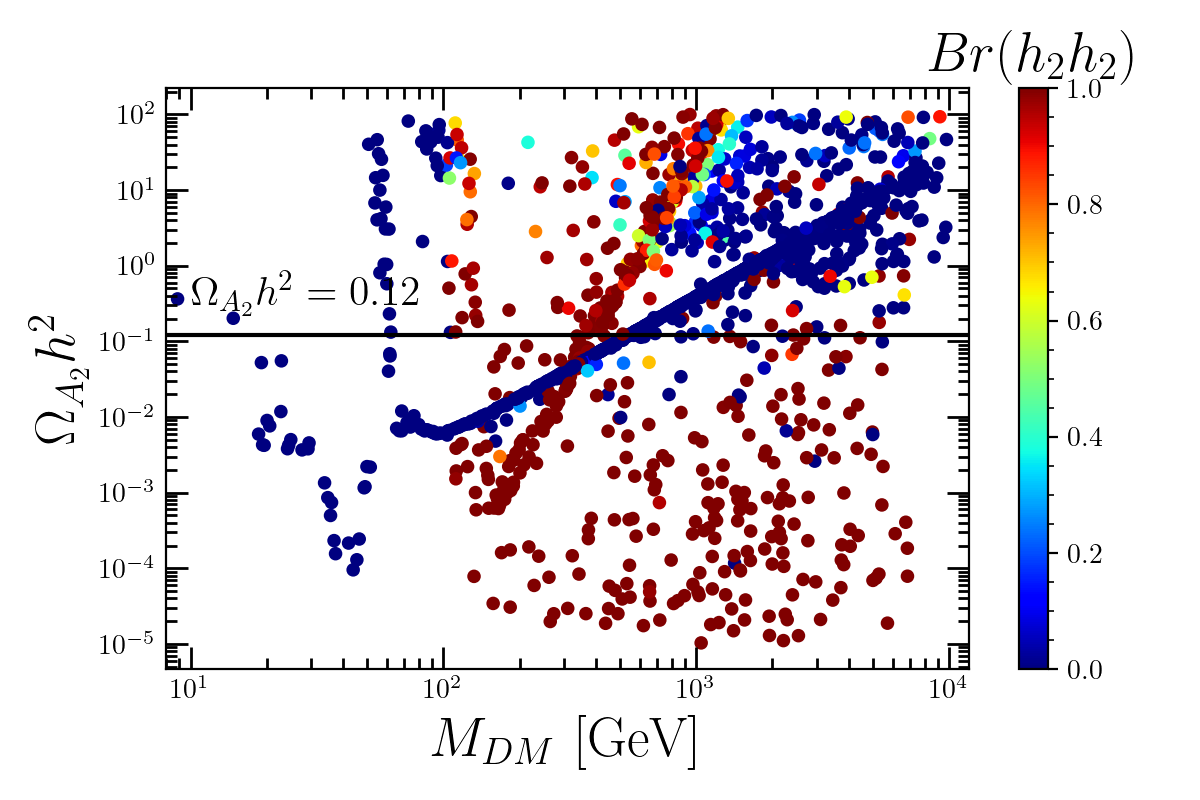}\label{Fig:8b}}
}
\caption{Scenario IV: singlet-like CP-odd scalar dark matter. (a) The relic density as a function of the $M_{DM}$ for a varying $|\lambda_{A_2 A_2 h_2}|$. The parameter space is consistent with the direct detection bound and other constraints. (b) The same as the left plot, but for a varying branching ratio of the dark matter annihilation to $h_2 h_2$.} 
\label{Fig:A2DM}
\end{figure}

In Fig.~\ref{Fig:8a}, we show the relic density as a function of dark matter mass for a varying $\lambda_{A_2 A_2 h_2}$. Here, all the points are allowed by the direct detection constraints and other above mentioned constraints. In Fig.~\ref{Fig:8a}, we show variation of branching ratio for the process $DM~DM \to h_2 h_2$ in $M_{DM}$ vs $\Omega_{A_2}h^2$ plane. In the parameter space with $\Omega h^2\le 0.12$, the dark matter annihilation is dominated by $A_2 A_2 \to h_2 h_2$, which makes up $>95\%$ contribution to the total annihilation cross section.  The contact term contribution to $A_2 A_2 \to h_2 h_2$ is suppressed for a small  value of $\lambda_{S \varphi}$, but the s-channel and t-channel contributions to the corresponding DM annihilation cross section can be sizable due to a sizable value of the trilinear coupling, $|\lambda_{A_2 A_2 h_2}|= |2\mu-\lambda_{S \varphi} v_\varphi|$.  For instance, for the choice of parameters in Fig.~\ref{Fig:8b}, $\mu=100\,{\rm GeV}$, $v_\varphi=10^4\,{\rm GeV}$ and $\lambda_{S\varphi}=[10^{-4} ,1]$, we can take $\lambda_{A_2A_2h_2}$ to be as large as $10^4\, {\rm GeV}$, so the contact term contribution to $A_2 A_2 \to h_2 h_2$ becomes subdominant for a small  value of $\lambda_{S \varphi}$.

\section{Summary and Conclusions} \label{sec:summary}

We have considered the extension of the SM with an inert scalar doublet, three RH neutrinos and several singlet scalar fields. In this model, neutrino masses are zero in the limit of the unbroken $Z_4$ discrete symmetry when the mixing angles between the neutral components of the inert scalar doublet and the singlet scalar $S$ vanish. Thus, small neutrino masses can be generated due to a nonzero VEV of the singlet scalar field $\varphi$, which induces the DM mixings and splits the DM scalars, whereas the lightest neutral scalar field with $Z_2$-odd parity can be a dark matter candidate. 

We consider various possibilities for neutrino masses and scalar dark matter in our model, depending on the DM mixing angles. 
In the scenario I where the DM mixings are zero, the neutral components of the inert scalar doublet have degenerate masses, so neutrino masses vanish. Thus, we need to rely on a different mechanism for neutrino masses such as high-scale seesaw. In this scenario, the inert scalar doublet could not be regarded as dark matter due to strong direct detection bounds. Instead, it is possible to regard the real scalar or complex scalar field coming from $S$ as being a dark matter candidate, which can constitute the whole of the observed dark matter through the Higgs-portal interaction and the extra DM annihilation into a pair of dark Higgs scalars (Fig. 2-(a)). 

In the scenario II where the DM mixing angles are small due to either small DM mixing masses or decoupled $S$, we can realize the lighter neutral component of the inert scalar doublet as a dark matter candidate and obtain small neutrino masses. For a sufficiently large mass splitting within the inert scalar doublet, we can evade the bounds from direct detection and the DM annihilation into a pair of dark Higgs scalars can help enlarge the parameter space for the correct relic density when the effective trilinear coupling between DM and the dark Higgs is sizable (Fig. 4-(a)). 

Furthermore, in the scenario III where the DM mixing angles are maximal, the DM candidate is a mixture of singlet and inert doublet scalars, so the Higgs-portal couplings for dark matter are constrained most by direct detection as well as electroweak precision data. When one of the DM mixing angles is small, neutrino masses can be suppressed by the square of the small DM mixing angle. On the other hand, when the DM mixing angles for both CP-even and CP-odd neutral scalars are maximal,  the mass splittings between the neutral DM scalars are constrained by electroweak precision data, so the resulting neutrino masses get suppressed too. 

The $Z_4$ breaking mass parameters, $\mu$ and $\kappa'$, split the masses for singlet and/or doublet scalars, being proportional to the VEV of the singlet scalar field $\varphi$. In the scenario III with bi-maximal DM mixings, the $\mu$ term is zero and the $\kappa'$ term is constrained to be small by direct detection, without a fine-tuning. Thus, the neutral DM scalars have almost degenerate masses, so there is a region with low DM masses that is consistent with the relic density, due to the efficient co-annihilation channels (Fig. 6-(b)). For large DM masses, the effective trilinear coupling between DM and the dark Higgs is crucial for determining the relic density through the DM annihilation channels into a pair of dark Higgs scalars, as in scenarios I and II (Fig. 6-(a) and (b)).
As a result, in the scenario III, both electroweak precision data and direct detection bounds favor almost degenerate masses for DM scalars, for which small neutrino masses are generated.

\section*{Acknowledgements}

The work is supported in part by Basic Science Research Program through the National
Research Foundation of Korea (NRF) funded by the Ministry of Education, Science and
Technology (NRF-2022R1A2C2003567). 
This research was supported by the Chung-Ang University Research
Grants in 2025.

\appendix 
\section{Dark matter interaction vertices}\label{scalar_coupling}
In this appendix, we list the gauge and scalar interactions for dark matter and its heavier partners.
Here, we denote the mixing angles for dark matter as $s_{\theta_s}=\sin\theta_s, s_{\theta_a}=\sin\theta_a$, etc, and the Higgs mixing angle as $s_\alpha=\sin\alpha$, etc. 

\vspace{0.5cm}
\noindent
{\bf Gauge interactions:}
{\small
\begin{align}
\mathcal{L} &\supset \frac{e^2}{8c^2_W s^2_W} \left( \left(A_1 c_{\theta_a} -A_2 s_{\theta_a} \right)^2 + \left( H_1 c_{\theta_s} -H_2 s_{\theta_s}  \right)^2 \right) \left( Z^\mu Z_\mu +2 c^2_W W^+_\mu W^{-\mu} \right) \nonumber\\
&\quad +\frac{e}{2 c_W s_W} \left( \left(   H_1 c_{\theta_s} -H_2 s_{\theta_s}  \right) \partial_\mu \left( A_1 c_{\theta_a} -A_2 s_{\theta_a} \right) +  \left(A_1 c_{\theta_a} -A_2 s_{\theta_a} \right) \partial_\mu \left(   H_1 c_{\theta_s} -H_2 s_{\theta_s} \right)    \right)Z^\mu  \nonumber\\
&\quad +i \frac{e}{2 s_W} \left( \left( c_{\theta_s} H_1 -s_{\theta_s} H_2  \right) \partial_\mu H^- - H^+ \partial \left( c_{\theta_s} H_1 -s_{\theta_s} H_2 \right)  \right)  W^-_\mu  \nonumber\\
&\quad +\frac{e}{2 s_W} \left( \left(A_1 c_{\theta_a} -A_2 s_{\theta_a} \right) \partial_\mu H^+ - H^+ \partial_\mu  \left(A_1 c_{\theta_a} -A_2 s_{\theta_a} \right)     \right)W^-_\mu + {\rm h.c.}
\end{align}
}

\noindent
{\bf Scalar quartic interactions: }
{\small
\begin{align}
-i\lambda_{A_1 A_1 h_1 h_1}&=-c_{\alpha }^2 \left(\left(\lambda _3+\lambda _4\right) c_{\theta _a}^2+\lambda _{H_1 S} s_{\theta _a}^2\right)-s_{\alpha }^2 \left(\lambda _{H_2 \varphi} c_{\theta _a}^2+\lambda _{S \varphi} s_{\theta _a}^2\right)+2 c_{\alpha } s_{\alpha } \frac{2\kappa^\prime}{v_\varphi} c_{\theta _a} s_{\theta _a}, \nonumber\\
-i\lambda_{A_2 A_1 h_1 h_1}&=c_{\alpha }^2 \left(-\lambda _{H_1 S}+\lambda _3+\lambda _4\right) c_{\theta _a} s_{\theta _a}+s_{\alpha }^2 c_{\theta _a} s_{\theta _a} \left(\lambda _{H_2 \varphi}-\lambda _{S \varphi}\right)+c_{\alpha } s_{\alpha } \frac{2\kappa^\prime}{v_\varphi} c_{2 \theta _a}, \nonumber\\
-i\lambda_{A_2,A_2,h_1,h_1}&=-c_{\alpha }^2 \left(\lambda _{H_1 S} c_{\theta _a}^2+\left(\lambda _3+\lambda _4\right) s_{\theta _a}^2\right)-s_{\alpha }^2 \left(\lambda _{S \varphi} c_{\theta _a}^2+\lambda _{H_2 \varphi} s_{\theta _a}^2\right)-2 c_{\alpha } s_{\alpha } \frac{2\kappa^\prime}{v_\varphi} c_{\theta _a} s_{\theta _a}, \nonumber\\
-i\lambda_{A_1 A_1 h_2 h_1}&=c_{\alpha } s_{\alpha } \left(\left(-\lambda _{H_2 \varphi}+\lambda _3+\lambda _4\right) c_{\theta _a}^2+s_{\theta _a}^2 \left(\lambda _{H_1 S}-\lambda _{S \varphi}\right)\right)+c_{\alpha }^2 \frac{2\kappa^\prime}{v_\varphi} c_{\theta _a} s_{\theta _a}-s_{\alpha }^2 \frac{2\kappa^\prime}{v_\varphi} c_{\theta _a} s_{\theta _a}, \nonumber\\
-2i\lambda_{A_2 A_1 h_2 h_1}&=-c_{\alpha } s_{\alpha } s_{2 \theta _a} \left(-\lambda _{H_1 S}-\lambda _{H_2 \varphi}+\lambda _3+\lambda _4+\lambda _{S \varphi}\right)-s_{\alpha }^2 \frac{2\kappa^\prime}{v_\varphi} c_{2 \theta _a}+c_{\alpha }^2 \frac{2\kappa^\prime}{v_\varphi} c_{2 \theta _a}, \nonumber\\
-i\lambda_{A_2 A_2 h_2 h_1}&=c_{\alpha } s_{\alpha } \left(c_{\theta _a}^2 \left(\lambda _{H_1 S}-\lambda _{S \varphi}\right)+\left(-\lambda _{H_2 \varphi}+\lambda _3+\lambda _4\right) s_{\theta _a}^2\right)+c_{\alpha }^2 \frac{2\kappa^\prime}{v_\varphi} \left(-c_{\theta _a}\right) s_{\theta _a}+s_{\alpha }^2 \frac{2\kappa^\prime}{v_\varphi} c_{\theta _a} s_{\theta _a}, \nonumber\\
-i\lambda_{A_1 A_1 h_2 h_2}&=-s_{\alpha }^2 \left(\left(\lambda _3+\lambda _4\right) c_{\theta _a}^2+\lambda _{H_1 S} s_{\theta _a}^2\right)-\left(c_{\alpha }^2 \left(\lambda _{H_2 \varphi} c_{\theta _a}^2+\lambda _{S \varphi} s_{\theta _a}^2\right)\right)-2 c_{\alpha } s_{\alpha } \frac{2\kappa^\prime}{v_\varphi} c_{\theta _a} s_{\theta _a}, \nonumber\\
-i\lambda_{A_2 A_1 h_2 h_2}&=s_{\alpha }^2 c_{\theta _a} s_{\theta _a} \left(-\lambda _{H_1 S}+\lambda _3+\lambda _4\right) +c_{\alpha }^2 c_{\theta _a} s_{\theta _a} \left(\lambda _{H_2 \varphi}-\lambda _{S \varphi}\right)-c_{\alpha } s_{\alpha } \frac{2\kappa^\prime}{v_\varphi} c_{2 \theta _a}, \nonumber\\
-i\lambda_{A_2 A_2 h_2 h_2}&=-s_{\alpha }^2 \left(\lambda _{H_1 S} c_{\theta _a}^2+\left(\lambda _3+\lambda _4\right) s_{\theta _a}^2\right)-c_{\alpha }^2 \left(\lambda _{S \varphi} c_{\theta _a}^2+\lambda _{H_2 \varphi} s_{\theta _a}^2\right)+2 c_{\alpha } s_{\alpha } \frac{2\kappa^\prime}{v_\varphi} c_{\theta _a} s_{\theta _a},
\end{align}
}
and
{\small
\begin{align}
-i\lambda_{H_1 H_1 h_1 h_1}&=-c_{\alpha }^2 \left(\left(\lambda _3+\lambda _4\right) c_{\theta _s}^2+\lambda _{H_1 S} s_{\theta _s}^2\right)-s_{\alpha }^2 \left(\lambda _{H_2 \varphi} c_{\theta _s}^2+s_{\theta _s}^2 \lambda _{S \varphi}\right)-2 c_{\alpha } s_{\alpha } s_{\theta _s} \frac{2\kappa^\prime}{v_\varphi} c_{\theta _s},\nonumber\\
-i\lambda_{H_2 H_1 h_1 h_1}&=c_{\alpha }^2 s_{\theta _s} c_{\theta _s} \left(-\lambda _{H_1 S}+\lambda _3+\lambda _4\right) +s_{\alpha }^2 s_{\theta _s} c_{\theta _s} \left(\lambda _{H_2 \varphi}-\lambda _{S \varphi}\right)-c_{\alpha } s_{\alpha } \frac{2\kappa^\prime}{v_\varphi} c_{2 \theta _s},\nonumber\\
-i\lambda_{H_2 H_2 h_1 h_1}&=-c_{\alpha }^2 \left(\lambda _{H_1 S} c_{\theta _s}^2+\left(\lambda _3+\lambda _4\right) s_{\theta _s}^2\right)-s_{\alpha }^2 \left(\lambda _{S \varphi} c_{\theta _s}^2+\lambda _{H_2 \varphi} s_{\theta _s}^2\right)+2 c_{\alpha } s_{\alpha } s_{\theta _s} \frac{2\kappa^\prime}{v_\varphi} c_{\theta _s},\nonumber\\
-i\lambda_{H_1 H_1 h_2 h_1}&=c_{\alpha } s_{\alpha } \left(\left(-\lambda _{H_2 \varphi}+\lambda _3+\lambda _4\right) c_{\theta _s}^2+s_{\theta _s}^2 \left(\lambda _{H_1 S}-\lambda _{S \varphi}\right)\right)-(c_{\alpha }^2-s_{\alpha }^2) s_{\theta _s} c_{\theta _s}\frac{2\kappa^\prime}{v_\varphi}, \nonumber\\
-2i\lambda_{H_2,H_1,h_2,h_1}&=-c_{\alpha } s_{\alpha } s_{2 \theta _s} \left(-\lambda _{H_1 S}-\lambda _{H_2 \varphi}+\lambda _3+\lambda _4+\lambda _{S \varphi}\right)-(c_{\alpha }^2 -s_{\alpha }^2) c_{2 \theta _s}\frac{2\kappa^\prime}{v_\varphi}, \nonumber\\
-i\lambda_{H_2 H_2 h_2 h_1}&= c_{\alpha } s_{\alpha } \left(c_{\theta _s}^2 \left(\lambda _{H_1 S}-\lambda _{S \varphi}\right)+\left(-\lambda _{H_2 \varphi}+\lambda _3+\lambda _4\right) s_{\theta _s}^2\right)+(c_{\alpha }^2 -s_{\alpha }^2) s_{\theta _s} \frac{2\kappa^\prime}{v_\varphi} c_{\theta _s},\nonumber\\
-i\lambda_{H_1 H_1 h_2 h_2}&= -s_{\alpha }^2 \left(\left(\lambda _3+\lambda _4\right) c_{\theta _s}^2+\lambda _{H_1 S} s_{\theta _s}^2\right)-c_{\alpha }^2 \left(\lambda _{H_2 \varphi} c_{\theta _s}^2+s_{\theta _s}^2 \lambda _{S \varphi}\right)+2 c_{\alpha } s_{\alpha } s_{\theta _s} \frac{2\kappa^\prime}{v_\varphi} c_{\theta _s},\nonumber\\
-i\lambda_{H_2,H_1,h_2,h_2}&=\left(-\lambda _{H_1 S}+\lambda _3+\lambda _4\right) s_{\alpha }^2 s_{\theta _s} c_{\theta _s}+c_{\alpha }^2 s_{\theta _s} c_{\theta _s} \left(\lambda _{H_2 \varphi}-\lambda _{S \varphi}\right)+c_{\alpha } s_{\alpha } \frac{2\kappa^\prime}{v_\varphi} c_{2 \theta _s},\nonumber\\
-i\lambda_{H_2 H_2 h_2 h_2}&=-s_{\alpha }^2 \left(\lambda _{H_1 S} c_{\theta _s}^2+\left(\lambda _3+\lambda _4\right) s_{\theta _s}^2\right)-c_{\alpha }^2 \left(\lambda _{S \varphi} c_{\theta _s}^2+\lambda _{H_2 \varphi} s_{\theta _s}^2\right)-2 c_{\alpha } s_{\alpha } s_{\theta _s} \frac{2\kappa^\prime}{v_\varphi} c_{\theta _s}.
\end{align}
}

\noindent
{\bf Scalar tri-linear interactions: }
{\small
\begin{align}
-i\lambda_{ H_1 H_1 h_1}&=-c_{\alpha } \bigg((\kappa +\kappa^\prime) s_{2\theta _s}+\left(\lambda _3+\lambda _4\right) v_Hc_{\theta _s}^2+v_H\lambda _{H_1 S} s_{\theta _s}^2\bigg),\nonumber\\
&\quad-s_{\alpha } \left(\lambda _{H_2 \varphi} v_{\varphi } c_{\theta _s}^2+v_Hs_{\theta_s} \frac{2\kappa^\prime}{v_\varphi} c_{\theta _s}+s_{\theta _s}^2 \left(2 \mu +\lambda _{S \varphi} v_{\varphi }\right)\right),\nonumber\\
-2i\lambda_{ H_1 H_2 h_1}&=-c_{\alpha } \bigg(2 (\kappa +\kappa^\prime) c_{2 \theta _s}-v_H\left(-\lambda _{H_1 S}+\lambda _3+\lambda _4\right) s_{2 \theta _s}\bigg) \nonumber\\
&\quad-s_{\alpha } \left(v_H\frac{2\kappa^\prime}{v_\varphi} c_{2 \theta _s}+s_{2 \theta _s} \left(v_{\varphi } \left(\lambda_{S \varphi}-\lambda _{H_2 \varphi}\right)+2 \mu \right)\right),\nonumber\\
-i\lambda_{ H_2 H_2 h_1}&=- c_{\theta _s}^2 \bigg(v_Hc_{\alpha } \lambda _{H_1 S}+s_{\alpha } \left(2 \mu +\lambda_{S \varphi} v_{\varphi }\right)\bigg)-s_{\theta _s}^2 \bigg(\left(\lambda _3+\lambda _4\right) v_Hc_{\alpha }+\lambda_{H_2 \varphi} s_{\alpha } v_{\varphi }\bigg),\nonumber\\
&\quad+s_{2\theta _s} \left( c_{\alpha } (\kappa +\kappa^\prime)+v_Hs_{\alpha } \frac{\kappa^\prime}{v_\varphi}\right),\nonumber\\
-i\lambda_{H_1 H_1 h_2}&=s_{\theta _s}^2 \bigg(v_H\lambda_{H_1 S} s_{\alpha }-c_{\alpha } \left(2 \mu +\lambda_{S \varphi} v_{\varphi }\right)\bigg)+c_{\theta _s}^2 \bigg(\left(\lambda _3+\lambda _4\right) v_H s_{\alpha }-c_{\alpha } \lambda _{H_2 \varphi} v_{\varphi }\bigg)\nonumber\\
&\quad+s_{2\theta _s} \left( (\kappa +\kappa^\prime) s_{\alpha }-v_Hc_{\alpha } \frac{\kappa^\prime}{v_\varphi}\right),\nonumber\\
-2i\lambda_{ H_1 H_2 h_2}&=-s_{2 \theta _s} \bigg(c_{\alpha } \left(v_{\varphi } \left(\lambda _{S \varphi}-\lambda _{H_2 \varphi}\right)+2 \mu \right)+v_H\left(-\lambda _{H_1 S}+\lambda _3+\lambda _4\right) s_{\alpha }\bigg)\nonumber\\
&\quad-c_{2 \theta _s} \left(v_Hc_{\alpha } \frac{2\kappa^\prime}{v_\varphi}-2 (\kappa +\kappa^\prime) s_{\alpha }\right),\nonumber\\
-i\lambda_{ H_2 H_2 h_2}&=s_{\alpha } \bigg(v_H\lambda _{H_1 S} c_{\theta _s}^2-(\kappa +\kappa^\prime) s_{2\theta _s}+\left(\lambda _3+\lambda _4\right) v_H s_{\theta _s}^2\bigg)\nonumber\\
&\quad-c_{\alpha } \left(c_{\theta _s}^2 \left(2 \mu +\lambda _{S \varphi} v_{\varphi }\right)-v_Hs_{\theta _s} \frac{2\kappa^\prime}{v_\varphi} c_{\theta _s}+\lambda _{H_2 \varphi} s_{\theta_s}^2 v_{\varphi }\right),
\end{align}
}
and 
{\small
\begin{align}
-i\lambda_{A_1 A_1 h_1 }&= s_{\alpha } \left(-\lambda _{H_2 \varphi} v_{\varphi } c_{\theta _a}^2+v_H\frac{2\kappa^\prime}{v_\varphi} c_{\theta _a} s_{\theta _a}+s_{\theta _a}^2 \left(2 \mu -\lambda _{S \varphi} v_{\varphi }\right)\right),\nonumber\\
&\quad-c_{\alpha } \left((\kappa -\kappa^\prime) s_{2\theta _a}+\left(\lambda _3+\lambda _4\right) v_Hc_{\theta _a}^2+v_H\lambda _{H_1 S} s_{\theta _a}^2\right),\nonumber\\
-2i\lambda_{ A_1 A_2 h_1 }&= c_{\alpha } \left((2 \kappa^\prime-2 \kappa ) c_{2 \theta _a}+v_H\left(-\lambda _{H_1 S}+\lambda _3+\lambda _4\right) s_{2 \theta _a}\right),\nonumber\\
&\quad+s_{\alpha } \left(v_H\frac{2\kappa^\prime}{v_\varphi} c_{2 \theta _a}+s_{2 \theta _a} \left(\lambda _{H_2 \varphi} v_{\varphi }+2 \mu -\lambda _{S \varphi} v_{\varphi }\right)\right),\nonumber\\
-i\lambda_{ A_2 A_2 h_1 }&= s_{\alpha } \left(-v_H\frac{2\kappa^\prime}{v_\varphi} c_{\theta _a} s_{\theta _a}+c_{\theta _a}^2 \left(2 \mu -\lambda _{S \varphi} v_{\varphi }\right)-\lambda _{H_2 \varphi} v_{\varphi } s_{\theta _a}^2\right),\nonumber\\
&\quad-c_{\alpha } \left(v_H\lambda _{H_1 S} c_{\theta _a}^2+(\kappa^\prime-\kappa ) s_{2\theta _a}+\left(\lambda _3+\lambda _4\right) v_Hs_{\theta _a}^2\right),\nonumber\\
-i\lambda_{A_1 A_1 h_2 }&= s_{\alpha } \left((\kappa - \kappa^\prime)  s_{2\theta _a}+\left(\lambda _3+\lambda _4\right) v_Hc_{\theta _a}^2+v_H\lambda _{H_1 S} s_{\theta _a}^2\right),\nonumber\\
&\quad+c_{\alpha } \left(-\lambda _{H_2 \varphi} v_{\varphi } c_{\theta _a}^2+v_H\frac{2\kappa^\prime}{v_\varphi} c_{\theta _a} s_{\theta _a}+s_{\theta _a}^2 \left(2 \mu -\lambda _{S \varphi} v_{\varphi }\right)\right),\nonumber\\
-2i\lambda_{ A_1 A_2 h_2 }&=s_{\alpha } \left((2 \kappa -2 \kappa^\prime) c_{2 \theta _a}-v_H\left(-\lambda _{H_1 S}+\lambda _3+\lambda _4\right) s_{2 \theta _a}\right),\nonumber\\
&\quad+c_{\alpha } \left(v_H\frac{2\kappa^\prime}{v_\varphi} c_{2 \theta _a}+s_{2 \theta _a} \left(\lambda _{H_2 \varphi} v_{\varphi }+2 \mu -\lambda _{S \varphi} v_{\varphi }\right)\right),\nonumber\\  
-i\lambda_{A_2 A_2 h_2 }&= s_{\alpha } \left(v_H\lambda _{H_1 S} c_{\theta _a}^2+( \kappa^\prime-\kappa )  s_{2\theta _a}+\left(\lambda _3+\lambda _4\right) v_Hs_{\theta _a}^2\right),\nonumber\\
&\quad+c_{\alpha } \left(-v_H\frac{2\kappa^\prime}{v_\varphi} c_{\theta _a} s_{\theta _a}+c_{\theta _a}^2 \left(2 \mu -\lambda _{S \varphi} v_{\varphi }\right)-\lambda _{H_2 \varphi} v_{\varphi } s_{\theta _a}^2\right).
\end{align}
}

\section{Perturbative unitarity bounds} \label{appendix:b}

In the basis of $ \mathcal{B}_1 = \left\{ H_1^- H_1^+,H_2^- H_2^+,\frac{A^2}{\sqrt{2}},\frac{h^2}{\sqrt{2}},\frac{A_0^2}{\sqrt{2}},\frac{H_0^2}{\sqrt{2}},\frac{s^2}{\sqrt{2}},\frac{a^2}{\sqrt{2}},\frac{\rho ^2}{\sqrt{2}},\frac{\eta ^2}{\sqrt{2}} \right\}$, the scattering matrix is given by
\begin{equation}
{\tiny
\mathcal{M}_1=
\begin{pmatrix} 
 4 \lambda_1 & \lambda_3+\lambda_4 & \sqrt{2} \lambda_1 & \sqrt{2} \lambda_1 & \frac{\lambda_3}{\sqrt{2}} & \frac{\lambda_3}{\sqrt{2}} & \frac{\lambda_{H_1S}}{\sqrt{2}} & \frac{\lambda_{H_1S}}{\sqrt{2}} & \frac{\lambda_{H_1 \varphi}}{\sqrt{2}} & \frac{\lambda_{H_1 \varphi}}{\sqrt{2}} \\
 \lambda_3+\lambda_4 & 4 \lambda_2 & \frac{\lambda_3}{\sqrt{2}} & \frac{\lambda_3}{\sqrt{2}} & \sqrt{2} \lambda_2 & \sqrt{2} \lambda_2 & \frac{\lambda_{H_2 S}}{\sqrt{2}} & \frac{\lambda_{H_2 S}}{\sqrt{2}} & \frac{\lambda_{H_2 \varphi}}{\sqrt{2}} & \frac{\lambda_{H_2 \varphi}}{\sqrt{2}} \\
 \sqrt{2} \lambda_1 & \frac{\lambda_3}{\sqrt{2}} & 3 \lambda_1 & \lambda_1 & \frac{1}{2}(\lambda_3+\lambda_4) & \frac{1}{2}(\lambda_3+\lambda_4) & \frac{\lambda_{H_1 S}}{2} & \frac{\lambda_{H_1 S}}{2} & \frac{\lambda_{H_1 \varphi}}{2} & \frac{\lambda_{H_1 \varphi}}{2} \\
 \sqrt{2} \lambda_1 & \frac{\lambda_3}{\sqrt{2}} & \lambda_1 & 3 \lambda_1 & \frac{1}{2}(\lambda_3+\lambda_4) & \frac{1}{2}(\lambda_3+\lambda_4) & \frac{\lambda_{H_1 S}}{2} & \frac{\lambda_{H_1 S}}{2} & \frac{\lambda_{H_1 \varphi}}{2} & \frac{\lambda_{H_1 \varphi}}{2} \\
 \frac{\lambda_3}{\sqrt{2}} & \sqrt{2} \lambda_2 & \frac{1}{2}(\lambda_3+\lambda_4) & \frac{1}{2}(\lambda_3+\lambda_4) & 3 \lambda_2 & \lambda_2 & \frac{\lambda_{H_2 S}}{2} & \frac{\lambda_{H_2 S}}{2} & \frac{\lambda_{H_2 \varphi}}{2} & \frac{\lambda_{H_2 \varphi}}{2} \\
 \frac{\lambda_3}{\sqrt{2}} & \sqrt{2} \lambda_2 & \frac{1}{2}(\lambda_3+\lambda_4) & \frac{1}{2}(\lambda_3+\lambda_4) & \lambda_2 & 3 \lambda_2 & \frac{\lambda_{H_2 S}}{2} & \frac{\lambda_{H_2 S}}{2} & \frac{\lambda_{H_2 \varphi}}{2} & \frac{\lambda_{H_2 \varphi}}{2} \\
 \frac{\lambda_{H_1S}}{\sqrt{2}} & \frac{\lambda_{H_2 S}}{\sqrt{2}} & \frac{\lambda_{H_1 S}}{2} & \frac{\lambda_{H_1 S}}{2} & \frac{\lambda_{H_2 S}}{2} & \frac{\lambda_{H_2 S}}{2} & 3 \lambda_s & \lambda_s & \frac{\lambda_{S\varphi}}{2} & \frac{\lambda_{S\varphi}}{2} \\
 \frac{\lambda_{H_1S}}{\sqrt{2}} & \frac{\lambda_{H_2 S}}{\sqrt{2}} & \frac{\lambda_{H_1 S}}{2} & \frac{\lambda_{H_1 S}}{2} & \frac{\lambda_{H_2 S}}{2} & \frac{\lambda_{H_2 S}}{2} & \lambda_s & 3 \lambda_s & \frac{\lambda_{S\varphi}}{2} & \frac{\lambda_{S\varphi}}{2} \\
 \frac{\lambda_{H_1 \varphi}}{\sqrt{2}} & \frac{\lambda_{H_2 \varphi}}{\sqrt{2}} & \frac{\lambda_{H_1 \varphi}}{2} & \frac{\lambda_{H_1 \varphi}}{2} & \frac{\lambda_{H_2 \varphi}}{2} & \frac{\lambda_{H_2 \varphi}}{2} & \frac{\lambda_{S\varphi}}{2} & \frac{\lambda_{S\varphi}}{2} & 3 \lambda_\varphi & \lambda_\varphi \\
 \frac{\lambda_{H_1 \varphi}}{\sqrt{2}} & \frac{\lambda_{H_2 \varphi}}{\sqrt{2}} & \frac{\lambda_{H_1 \varphi}}{2} & \frac{\lambda_{H_1 \varphi}}{2} & \frac{\lambda_{H_2 \varphi}}{2} & \frac{\lambda_{H_2 \varphi}}{2} & \frac{\lambda_{S\varphi}}{2} & \frac{\lambda_{S\varphi}}{2} & \lambda_\varphi & 3 \lambda_\varphi
 \end{pmatrix},
 }
\end{equation}
so the corresponding eigenvalues are
  $\mathcal{E}_1=\left\{2 \lambda_1,2 \lambda_2,,2 \lambda_s,2 \lambda_\varphi,\lambda_1+\lambda_2\pm\sqrt{(\lambda_1- \lambda_2)^2+\lambda_4^2}\right\}$

In the basis of $\mathcal{B}_2=\left\{s H_1^+,s H_2^+,a H_1^+,a H_2^+,\rho  H_1^+,\rho  H_2^+,\eta  H_1^+,\eta  H_2^+\right\}$,
 the scattering matrix is
\begin{equation}
\mathcal{M}_2= \left(
\begin{array}{cccccccc}
 \lambda_{H_1 S} & 0 & 0 & 0 & 0 & \frac{\lambda_{S\varphi}^\prime}{2} & 0 & \frac{i \lambda_{S\varphi}^\prime}{2} \\
 0 & \lambda_{H_2 S} & 0 & 0 & \frac{\lambda_{S\varphi}^\prime}{2} & 0 & \frac{-i \lambda_{S\varphi}^\prime}{2}& 0 \\
 0 & 0 & \lambda_{H_1 S} & 0 & 0 & \frac{-i \lambda_{S\varphi}^\prime}{2}& 0 & \frac{\lambda_{S\varphi}^\prime}{2} \\
 0 & 0 & 0 & \lambda_{H_2 S} & \frac{i \lambda_{S\varphi}^\prime}{2} & 0 & \frac{\lambda_{S\varphi}^\prime}{2} & 0 \\
 0 & \frac{\lambda_{S\varphi}^\prime}{2} & 0 & \frac{-i \lambda_{S\varphi}^\prime}{2}& \lambda_{H_1 \varphi} & 0 & 0 & 0 \\
 \frac{\lambda_{S\varphi}^\prime}{2} & 0 & \frac{i \lambda_{S\varphi}^\prime}{2} & 0 & 0 & \lambda_{H_2 \varphi} & 0 & 0 \\
 0 & \frac{i \lambda_{S\varphi}^\prime}{2} & 0 & \frac{\lambda_{S\varphi}^\prime}{2} & 0 & 0 & \lambda_{H_1 \varphi} & 0 \\
 \frac{-i \lambda_{S\varphi}^\prime}{2}& 0 & \frac{\lambda_{S\varphi}^\prime}{2} & 0 & 0 & 0 & 0 & \lambda_{H_2 \varphi} \\
\end{array}
\right),
\end{equation}
so  the corresponding eigenvalues are
\begin{align}    
\mathcal{E}_2=&\bigg\{\lambda_{H_1 S},\lambda_{H_1 \varphi},\lambda_{H_2 S},\lambda_{H_2 \varphi},\frac{1}{2} \left(\lambda_{H_1 \varphi}+\lambda_{H_2 S}\pm\sqrt{(\lambda_{H_1 \varphi} -\lambda_{H_2 S})^2+4 \lambda_{S\varphi}^{\prime 2}}\right), \nonumber\\
&\quad\frac{1}{2} \left(\lambda_{H_1 S}+\lambda_{H_2 \varphi}\pm\sqrt{(\lambda_{H_1 S} -\lambda_{H_2 \varphi})^2+4 \lambda_{S\varphi}^{\prime 2}}\right)\bigg\}.
\end{align}

In the basis of $ \mathcal{B}_3=\left\{A h,A_0 H_0,a s,\eta  \rho\right\}$,
 the scattering matrix is
\begin{equation}
\mathcal{M}_3= 
\left(
\begin{array}{cccc}
 2 \lambda_{H_1 S} & 0 & 0 & 0 \\
 0 & 2 \lambda_{H_2 S} & 0 & 0 \\
 0 & 0 & 2 \lambda_{H_1 S} & 0 \\
 0 & 0 & 0 & 2 \lambda_{H_2 S} \\
\end{array}
\right), 
\end{equation}
so  the corresponding eigenvalues are
\bea
\mathcal{E}_3=\left\{ 2 \lambda_{H_1 S},2 \lambda_{H_2 S} \right\}.
\eea

In the basis of
$\mathcal{B}_4= \left\{H_1^+ H_2^-,H_1^- H_2^+,A_0 h,A H_0,A A_0,h H_0 \right\}$,
 the scattering matrix is
\begin{equation}
\mathcal{M}_4= \left(
\begin{array}{cccccc}
 0 & \lambda_3+\lambda_4 & \frac{i \lambda_4}{2} & -\frac{i \lambda_4}{2}  & \frac{\lambda_4}{2} & \frac{\lambda_4}{2} \\
 \lambda_3+\lambda_4 & 0 & -\frac{i \lambda_4}{2}  & \frac{i \lambda_4}{2} & \frac{\lambda_4}{2} & \frac{\lambda_4}{2} \\
 \frac{i \lambda_4}{2} & -\frac{i \lambda_4}{2}  & \lambda_3+\lambda_4 & 0 & 0 & 0 \\
 -\frac{i \lambda_4}{2}  & \frac{i \lambda_4}{2} & 0 & \lambda_3+\lambda_4 & 0 & 0 \\
 \frac{\lambda_4}{2} & \frac{\lambda_4}{2} & 0 & 0 & \lambda_3+\lambda_4 & 0 \\
 \frac{\lambda_4}{2} & \frac{\lambda_4}{2} & 0 & 0 & 0 & \lambda_3+\lambda_4 \\
\end{array}
\right),
\end{equation}
so  the corresponding eigenvalues are
\bea
\mathcal{E}_4=\left\{\lambda_3,\lambda_3+\lambda_4,\lambda_3+\lambda_4,\lambda_3+2 \lambda_4,\pm\sqrt{\lambda_3^2+2 \lambda_3 \lambda_4}\right\}.
\eea

Finally, in the basis of 
$\mathcal{B}_5=\left\{h H_1^+,H_0 H_1^+,A H_1^+,A_0 H_1^+,h H_2^+,H_0 H_2^+,A H_2^+,A_0 H_2^+ \right\}$,
the scattering matrix is
\begin{equation}
\mathcal{M}_5=\left(
\begin{array}{cccccccc}
 2 \lambda_1 & 0 & 0 & 0 & 0 & \frac{\lambda_4}{2} & 0 & \frac{i \lambda_4}{2} \\
 0 & \lambda_3 & 0 & 0 & \frac{\lambda_4}{2} & 0 & -\frac{i \lambda_4}{2}  & 0 \\
 0 & 0 & 2 \lambda_1 & 0 & 0 & -\frac{i \lambda_4}{2}  & 0 & \frac{\lambda_4}{2} \\
 0 & 0 & 0 & \lambda_3 & \frac{i \lambda_4}{2} & 0 & \frac{\lambda_4}{2} & 0 \\
 0 & \frac{\lambda_4}{2} & 0 & -\frac{i \lambda_4}{2}  & \lambda_3 & 0 & 0 & 0 \\
 \frac{\lambda_4}{2} & 0 & \frac{i \lambda_4}{2} & 0 & 0 & 2 \lambda_2 & 0 & 0 \\
 0 & \frac{i \lambda_4}{2} & 0 & \frac{\lambda_4}{2} & 0 & 0 & \lambda_3 & 0 \\
 -\frac{i \lambda_4}{2}  & 0 & \frac{\lambda_4}{2} & 0 & 0 & 0 & 0 & 2 \lambda_2 \\
\end{array}
\right), 
\end{equation}
so the corresponding eigenvalues are
\bea
\mathcal{E}_5=\left\{2 \lambda_1,2 \lambda_2,\lambda_3,\lambda_3-\lambda_4,\lambda_3+\lambda_4,\lambda_1+\lambda_2\pm\sqrt{(\lambda_1-\lambda_2)^2+\lambda_4^2} \right\}.
\eea

As a result, the perturbativity unitarity bounds set the eigenvalues of each scattering matrix to be smaller than $8\pi$.

\def\theequation{C.\arabic{equation}}

\setcounter{equation}{0}
\vskip0.8cm
\noindent
\section{Loop functions} \label{appendix:c}

The loop functions for the calculation of the oblique parameters are 
\begin{equation}
  F(I, J) = \frac{I + J}{2} - \frac{I J}{I - J} \ln{\frac{I}{J}},
\end{equation}
with $F(I,I) = 0$ in the limit of $J \to I$, and
\begin{equation}
\begin{split}
  G \left( I, J, Q \right) &=
  - \frac{16}{3} + \frac{5 \left( I + J \right)}{Q}
  - \frac{2 \left( I - J \right)^2}{Q^2}
  \\
  &+ \frac{3}{Q} \left[ \frac{I^2 + J^2}{I - J}
- \frac{I^2 - J^2}{Q}
+ \frac{\left( I - J \right)^3}{3 Q^2} \right]
\ln{\frac{I}{J}}
+ \frac{r}{Q^3}\, f \left( t, r \right),
\end{split}
\end{equation}
where
\begin{equation}
  t \equiv I + J - Q 
  \quad \text{and} \quad
  r \equiv Q^2 - 2 Q (I + J) + (I - J)^2,
\end{equation}
\begin{equation}
\begin{split}
  f \left( t, r \right) \equiv \left\{ \begin{array}{lcl}
{
\sqrt{r}\, \ln{\left| \frac{t - \sqrt{r}}{t + \sqrt{r}} \right|}
} & \Leftarrow r > 0,
\\*[3mm]
0 & \Leftarrow r = 0,
\\*[2mm]
{
2\, \sqrt{-r}\, \arctan{\frac{\sqrt{-r}}{t}}
} & \Leftarrow  r < 0.
\end{array} \right.
\end{split}
\end{equation}
\medskip

\clearpage
\bibliographystyle{apsrev4-1}
\bibliography{ref}

\begin{thebibliography}{22}%
\makeatletter
\providecommand \@ifxundefined [1]{%
 \@ifx{#1\undefined}
}%
\providecommand \@ifnum [1]{%
 \ifnum #1\expandafter \@firstoftwo
 \else \expandafter \@secondoftwo
 \fi
}%
\providecommand \@ifx [1]{%
 \ifx #1\expandafter \@firstoftwo
 \else \expandafter \@secondoftwo
 \fi
}%
\providecommand \natexlab [1]{#1}%
\providecommand \enquote  [1]{``#1''}%
\providecommand \bibnamefont  [1]{#1}%
\providecommand \bibfnamefont [1]{#1}%
\providecommand \citenamefont [1]{#1}%
\providecommand \href@noop [0]{\@secondoftwo}%
\providecommand \href [0]{\begingroup \@sanitize@url \@href}%
\providecommand \@href[1]{\@@startlink{#1}\@@href}%
\providecommand \@@href[1]{\endgroup#1\@@endlink}%
\providecommand \@sanitize@url [0]{\catcode `\\12\catcode `\$12\catcode
  `\&12\catcode `\#12\catcode `\^12\catcode `\_12\catcode `\%12\relax}%
\providecommand \@@startlink[1]{}%
\providecommand \@@endlink[0]{}%
\providecommand \url  [0]{\begingroup\@sanitize@url \@url }%
\providecommand \@url [1]{\endgroup\@href {#1}{\urlprefix }}%
\providecommand \urlprefix  [0]{URL }%
\providecommand \Eprint [0]{\href }%
\providecommand \doibase [0]{http://dx.doi.org/}%
\providecommand \selectlanguage [0]{\@gobble}%
\providecommand \bibinfo  [0]{\@secondoftwo}%
\providecommand \bibfield  [0]{\@secondoftwo}%
\providecommand \translation [1]{[#1]}%
\providecommand \BibitemOpen [0]{}%
\providecommand \bibitemStop [0]{}%
\providecommand \bibitemNoStop [0]{.\EOS\space}%
\providecommand \EOS [0]{\spacefactor3000\relax}%
\providecommand \BibitemShut  [1]{\csname bibitem#1\endcsname}%
\let\auto@bib@innerbib\@empty
\bibitem [{\citenamefont {Ma}(2006)}]{Ma:2006km}%
  \BibitemOpen
  \bibfield  {author} {\bibinfo {author} {\bibfnamefont {E.}~\bibnamefont
  {Ma}},\ }\href {\doibase 10.1103/PhysRevD.73.077301} {\bibfield  {journal}
  {\bibinfo  {journal} {Phys. Rev. D}\ }\textbf {\bibinfo {volume} {73}},\
  \bibinfo {pages} {077301} (\bibinfo {year} {2006})},\ \Eprint
  {http://arxiv.org/abs/hep-ph/0601225} {arXiv:hep-ph/0601225} \BibitemShut
  {NoStop}%
\bibitem [{\citenamefont {Minkowski}(1977)}]{Minkowski:1977sc}%
  \BibitemOpen
  \bibfield  {author} {\bibinfo {author} {\bibfnamefont {P.}~\bibnamefont
  {Minkowski}},\ }\href {\doibase 10.1016/0370-2693(77)90435-X} {\bibfield
  {journal} {\bibinfo  {journal} {Phys. Lett. B}\ }\textbf {\bibinfo {volume}
  {67}},\ \bibinfo {pages} {421} (\bibinfo {year} {1977})}\BibitemShut
  {NoStop}%
\bibitem [{\citenamefont {Yanagida}(1979)}]{Yanagida:1979as}%
  \BibitemOpen
  \bibfield  {author} {\bibinfo {author} {\bibfnamefont {T.}~\bibnamefont
  {Yanagida}},\ }\href@noop {} {\bibfield  {journal} {\bibinfo  {journal}
  {Conf. Proc. C}\ }\textbf {\bibinfo {volume} {7902131}},\ \bibinfo {pages}
  {95} (\bibinfo {year} {1979})}\BibitemShut {NoStop}%
\bibitem [{\citenamefont {Gell-Mann}\ \emph {et~al.}(1979)\citenamefont
  {Gell-Mann}, \citenamefont {Ramond},\ and\ \citenamefont
  {Slansky}}]{Gell-Mann:1979vob}%
  \BibitemOpen
  \bibfield  {author} {\bibinfo {author} {\bibfnamefont {M.}~\bibnamefont
  {Gell-Mann}}, \bibinfo {author} {\bibfnamefont {P.}~\bibnamefont {Ramond}}, \
  and\ \bibinfo {author} {\bibfnamefont {R.}~\bibnamefont {Slansky}},\
  }\href@noop {} {\bibfield  {journal} {\bibinfo  {journal} {Conf. Proc. C}\
  }\textbf {\bibinfo {volume} {790927}},\ \bibinfo {pages} {315} (\bibinfo
  {year} {1979})},\ \Eprint {http://arxiv.org/abs/1306.4669} {arXiv:1306.4669
  [hep-th]} \BibitemShut {NoStop}%
\bibitem [{\citenamefont {Mohapatra}\ and\ \citenamefont
  {Senjanovic}(1980)}]{Mohapatra:1979ia}%
  \BibitemOpen
  \bibfield  {author} {\bibinfo {author} {\bibfnamefont {R.~N.}\ \bibnamefont
  {Mohapatra}}\ and\ \bibinfo {author} {\bibfnamefont {G.}~\bibnamefont
  {Senjanovic}},\ }\href {\doibase 10.1103/PhysRevLett.44.912} {\bibfield
  {journal} {\bibinfo  {journal} {Phys. Rev. Lett.}\ }\textbf {\bibinfo
  {volume} {44}},\ \bibinfo {pages} {912} (\bibinfo {year} {1980})}\BibitemShut
  {NoStop}%
\bibitem [{\citenamefont {Schechter}\ and\ \citenamefont
  {Valle}(1980)}]{Schechter:1980gr}%
  \BibitemOpen
  \bibfield  {author} {\bibinfo {author} {\bibfnamefont {J.}~\bibnamefont
  {Schechter}}\ and\ \bibinfo {author} {\bibfnamefont {J.~W.~F.}\ \bibnamefont
  {Valle}},\ }\href {\doibase 10.1103/PhysRevD.22.2227} {\bibfield  {journal}
  {\bibinfo  {journal} {Phys. Rev. D}\ }\textbf {\bibinfo {volume} {22}},\
  \bibinfo {pages} {2227} (\bibinfo {year} {1980})}\BibitemShut {NoStop}%
\bibitem [{\citenamefont {Kim}\ \emph {et~al.}(2025)\citenamefont {Kim},
  \citenamefont {Kim}, \citenamefont {Lee},\ and\ \citenamefont
  {Padhan}}]{Kim:2024cwp}%
  \BibitemOpen
  \bibfield  {author} {\bibinfo {author} {\bibfnamefont {J.}~\bibnamefont
  {Kim}}, \bibinfo {author} {\bibfnamefont {S.-S.}\ \bibnamefont {Kim}},
  \bibinfo {author} {\bibfnamefont {H.~M.}\ \bibnamefont {Lee}}, \ and\
  \bibinfo {author} {\bibfnamefont {R.}~\bibnamefont {Padhan}},\ }\href
  {\doibase 10.1016/j.physletb.2025.139243} {\bibfield  {journal} {\bibinfo
  {journal} {Phys. Lett. B}\ }\textbf {\bibinfo {volume} {861}},\ \bibinfo
  {pages} {139243} (\bibinfo {year} {2025})},\ \Eprint
  {http://arxiv.org/abs/2407.13595} {arXiv:2407.13595 [hep-ph]} \BibitemShut
  {NoStop}%
\bibitem [{\citenamefont {de~la Vega}\ \emph {et~al.}(2024)\citenamefont {de~la
  Vega}, \citenamefont {Fitzpatrick}, \citenamefont {Martinez-Ramirez},\ and\
  \citenamefont {Peinado}}]{delaVega:2024tuu}%
  \BibitemOpen
  \bibfield  {author} {\bibinfo {author} {\bibfnamefont {L.~M.~G.}\
  \bibnamefont {de~la Vega}}, \bibinfo {author} {\bibfnamefont {P.~J.}\
  \bibnamefont {Fitzpatrick}}, \bibinfo {author} {\bibfnamefont
  {R.}~\bibnamefont {Martinez-Ramirez}}, \ and\ \bibinfo {author}
  {\bibfnamefont {E.}~\bibnamefont {Peinado}},\ }\href {\doibase
  10.1103/PhysRevD.110.115024} {\bibfield  {journal} {\bibinfo  {journal}
  {Phys. Rev. D}\ }\textbf {\bibinfo {volume} {110}},\ \bibinfo {pages}
  {115024} (\bibinfo {year} {2024})},\ \Eprint
  {http://arxiv.org/abs/2407.14447} {arXiv:2407.14447 [hep-ph]} \BibitemShut
  {NoStop}%
\bibitem [{\citenamefont {Belanger}\ \emph {et~al.}(2022)\citenamefont
  {Belanger}, \citenamefont {Mjallal},\ and\ \citenamefont
  {Pukhov}}]{Belanger:2021lwd}%
  \BibitemOpen
  \bibfield  {author} {\bibinfo {author} {\bibfnamefont {G.}~\bibnamefont
  {Belanger}}, \bibinfo {author} {\bibfnamefont {A.}~\bibnamefont {Mjallal}}, \
  and\ \bibinfo {author} {\bibfnamefont {A.}~\bibnamefont {Pukhov}},\ }\href
  {\doibase 10.1103/PhysRevD.105.035018} {\bibfield  {journal} {\bibinfo
  {journal} {Phys. Rev. D}\ }\textbf {\bibinfo {volume} {105}},\ \bibinfo
  {pages} {035018} (\bibinfo {year} {2022})},\ \Eprint
  {http://arxiv.org/abs/2108.08061} {arXiv:2108.08061 [hep-ph]} \BibitemShut
  {NoStop}%
\bibitem [{\citenamefont {Arcadi}\ \emph {et~al.}(2024)\citenamefont {Arcadi},
  \citenamefont {Cabo-Almeida}, \citenamefont {Dutra}, \citenamefont {Ghosh},
  \citenamefont {Lindner}, \citenamefont {Mambrini}, \citenamefont {Neto},
  \citenamefont {Pierre}, \citenamefont {Profumo},\ and\ \citenamefont
  {Queiroz}}]{Arcadi:2024ukq}%
  \BibitemOpen
  \bibfield  {author} {\bibinfo {author} {\bibfnamefont {G.}~\bibnamefont
  {Arcadi}}, \bibinfo {author} {\bibfnamefont {D.}~\bibnamefont
  {Cabo-Almeida}}, \bibinfo {author} {\bibfnamefont {M.}~\bibnamefont {Dutra}},
  \bibinfo {author} {\bibfnamefont {P.}~\bibnamefont {Ghosh}}, \bibinfo
  {author} {\bibfnamefont {M.}~\bibnamefont {Lindner}}, \bibinfo {author}
  {\bibfnamefont {Y.}~\bibnamefont {Mambrini}}, \bibinfo {author}
  {\bibfnamefont {J.~P.}\ \bibnamefont {Neto}}, \bibinfo {author}
  {\bibfnamefont {M.}~\bibnamefont {Pierre}}, \bibinfo {author} {\bibfnamefont
  {S.}~\bibnamefont {Profumo}}, \ and\ \bibinfo {author} {\bibfnamefont
  {F.~S.}\ \bibnamefont {Queiroz}},\ }\href@noop {} {\  (\bibinfo {year}
  {2024})},\ \Eprint {http://arxiv.org/abs/2403.15860} {arXiv:2403.15860
  [hep-ph]} \BibitemShut {NoStop}%
\bibitem [{\citenamefont {Gong}\ \emph {et~al.}(2012)\citenamefont {Gong},
  \citenamefont {Lee},\ and\ \citenamefont {Kang}}]{Gong:2012ri}%
  \BibitemOpen
  \bibfield  {author} {\bibinfo {author} {\bibfnamefont {J.-O.}\ \bibnamefont
  {Gong}}, \bibinfo {author} {\bibfnamefont {H.~M.}\ \bibnamefont {Lee}}, \
  and\ \bibinfo {author} {\bibfnamefont {S.~K.}\ \bibnamefont {Kang}},\ }\href
  {\doibase 10.1007/JHEP04(2012)128} {\bibfield  {journal} {\bibinfo  {journal}
  {JHEP}\ }\textbf {\bibinfo {volume} {04}},\ \bibinfo {pages} {128} (\bibinfo
  {year} {2012})},\ \Eprint {http://arxiv.org/abs/1202.0288} {arXiv:1202.0288
  [hep-ph]} \BibitemShut {NoStop}%
\bibitem [{\citenamefont {Lee}\ \emph {et~al.}(2011{\natexlab{a}})\citenamefont
  {Lee}, \citenamefont {Raby}, \citenamefont {Ratz}, \citenamefont {Ross},
  \citenamefont {Schieren}, \citenamefont {Schmidt-Hoberg},\ and\ \citenamefont
  {Vaudrevange}}]{Lee:2010gv}%
  \BibitemOpen
  \bibfield  {author} {\bibinfo {author} {\bibfnamefont {H.~M.}\ \bibnamefont
  {Lee}}, \bibinfo {author} {\bibfnamefont {S.}~\bibnamefont {Raby}}, \bibinfo
  {author} {\bibfnamefont {M.}~\bibnamefont {Ratz}}, \bibinfo {author}
  {\bibfnamefont {G.~G.}\ \bibnamefont {Ross}}, \bibinfo {author}
  {\bibfnamefont {R.}~\bibnamefont {Schieren}}, \bibinfo {author}
  {\bibfnamefont {K.}~\bibnamefont {Schmidt-Hoberg}}, \ and\ \bibinfo {author}
  {\bibfnamefont {P.~K.~S.}\ \bibnamefont {Vaudrevange}},\ }\href {\doibase
  10.1016/j.physletb.2010.10.038} {\bibfield  {journal} {\bibinfo  {journal}
  {Phys. Lett. B}\ }\textbf {\bibinfo {volume} {694}},\ \bibinfo {pages} {491}
  (\bibinfo {year} {2011}{\natexlab{a}})},\ \Eprint
  {http://arxiv.org/abs/1009.0905} {arXiv:1009.0905 [hep-ph]} \BibitemShut
  {NoStop}%
\bibitem [{\citenamefont {Lee}\ \emph {et~al.}(2011{\natexlab{b}})\citenamefont
  {Lee}, \citenamefont {Raby}, \citenamefont {Ratz}, \citenamefont {Ross},
  \citenamefont {Schieren}, \citenamefont {Schmidt-Hoberg},\ and\ \citenamefont
  {Vaudrevange}}]{Lee:2011dya}%
  \BibitemOpen
  \bibfield  {author} {\bibinfo {author} {\bibfnamefont {H.~M.}\ \bibnamefont
  {Lee}}, \bibinfo {author} {\bibfnamefont {S.}~\bibnamefont {Raby}}, \bibinfo
  {author} {\bibfnamefont {M.}~\bibnamefont {Ratz}}, \bibinfo {author}
  {\bibfnamefont {G.~G.}\ \bibnamefont {Ross}}, \bibinfo {author}
  {\bibfnamefont {R.}~\bibnamefont {Schieren}}, \bibinfo {author}
  {\bibfnamefont {K.}~\bibnamefont {Schmidt-Hoberg}}, \ and\ \bibinfo {author}
  {\bibfnamefont {P.~K.~S.}\ \bibnamefont {Vaudrevange}},\ }\href {\doibase
  10.1016/j.nuclphysb.2011.04.009} {\bibfield  {journal} {\bibinfo  {journal}
  {Nucl. Phys. B}\ }\textbf {\bibinfo {volume} {850}},\ \bibinfo {pages} {1}
  (\bibinfo {year} {2011}{\natexlab{b}})},\ \Eprint
  {http://arxiv.org/abs/1102.3595} {arXiv:1102.3595 [hep-ph]} \BibitemShut
  {NoStop}%
\bibitem [{\citenamefont {Lebedev}\ and\ \citenamefont
  {Lee}(2011)}]{Lebedev:2011aq}%
  \BibitemOpen
  \bibfield  {author} {\bibinfo {author} {\bibfnamefont {O.}~\bibnamefont
  {Lebedev}}\ and\ \bibinfo {author} {\bibfnamefont {H.~M.}\ \bibnamefont
  {Lee}},\ }\href {\doibase 10.1140/epjc/s10052-011-1821-0} {\bibfield
  {journal} {\bibinfo  {journal} {Eur. Phys. J. C}\ }\textbf {\bibinfo {volume}
  {71}},\ \bibinfo {pages} {1821} (\bibinfo {year} {2011})},\ \Eprint
  {http://arxiv.org/abs/1105.2284} {arXiv:1105.2284 [hep-ph]} \BibitemShut
  {NoStop}%
\bibitem [{\citenamefont {Navas}\ \emph {et~al.}(2024)\citenamefont {Navas}
  \emph {et~al.}}]{ParticleDataGroup:2024cfk}%
  \BibitemOpen
  \bibfield  {author} {\bibinfo {author} {\bibfnamefont {S.}~\bibnamefont
  {Navas}} \emph {et~al.} (\bibinfo {collaboration} {Particle Data Group}),\
  }\href {\doibase 10.1103/PhysRevD.110.030001} {\bibfield  {journal} {\bibinfo
   {journal} {Phys. Rev. D}\ }\textbf {\bibinfo {volume} {110}},\ \bibinfo
  {pages} {030001} (\bibinfo {year} {2024})}\BibitemShut {NoStop}%
\bibitem [{\citenamefont {Grimus}\ \emph {et~al.}(2008)\citenamefont {Grimus},
  \citenamefont {Lavoura}, \citenamefont {Ogreid},\ and\ \citenamefont
  {Osland}}]{Grimus:2008nb}%
  \BibitemOpen
  \bibfield  {author} {\bibinfo {author} {\bibfnamefont {W.}~\bibnamefont
  {Grimus}}, \bibinfo {author} {\bibfnamefont {L.}~\bibnamefont {Lavoura}},
  \bibinfo {author} {\bibfnamefont {O.~M.}\ \bibnamefont {Ogreid}}, \ and\
  \bibinfo {author} {\bibfnamefont {P.}~\bibnamefont {Osland}},\ }\href
  {\doibase 10.1016/j.nuclphysb.2008.04.019} {\bibfield  {journal} {\bibinfo
  {journal} {Nucl. Phys. B}\ }\textbf {\bibinfo {volume} {801}},\ \bibinfo
  {pages} {81} (\bibinfo {year} {2008})},\ \Eprint
  {http://arxiv.org/abs/0802.4353} {arXiv:0802.4353 [hep-ph]} \BibitemShut
  {NoStop}%
\bibitem [{\citenamefont {Pierce}\ and\ \citenamefont
  {Thaler}(2007)}]{Pierce:2007ut}%
  \BibitemOpen
  \bibfield  {author} {\bibinfo {author} {\bibfnamefont {A.}~\bibnamefont
  {Pierce}}\ and\ \bibinfo {author} {\bibfnamefont {J.}~\bibnamefont
  {Thaler}},\ }\href {\doibase 10.1088/1126-6708/2007/08/026} {\bibfield
  {journal} {\bibinfo  {journal} {JHEP}\ }\textbf {\bibinfo {volume} {08}},\
  \bibinfo {pages} {026} (\bibinfo {year} {2007})},\ \Eprint
  {http://arxiv.org/abs/hep-ph/0703056} {arXiv:hep-ph/0703056} \BibitemShut
  {NoStop}%
\bibitem [{\citenamefont {Lundstrom}\ \emph {et~al.}(2009)\citenamefont
  {Lundstrom}, \citenamefont {Gustafsson},\ and\ \citenamefont
  {Edsjo}}]{Lundstrom:2008ai}%
  \BibitemOpen
  \bibfield  {author} {\bibinfo {author} {\bibfnamefont {E.}~\bibnamefont
  {Lundstrom}}, \bibinfo {author} {\bibfnamefont {M.}~\bibnamefont
  {Gustafsson}}, \ and\ \bibinfo {author} {\bibfnamefont {J.}~\bibnamefont
  {Edsjo}},\ }\href {\doibase 10.1103/PhysRevD.79.035013} {\bibfield  {journal}
  {\bibinfo  {journal} {Phys. Rev. D}\ }\textbf {\bibinfo {volume} {79}},\
  \bibinfo {pages} {035013} (\bibinfo {year} {2009})},\ \Eprint
  {http://arxiv.org/abs/0810.3924} {arXiv:0810.3924 [hep-ph]} \BibitemShut
  {NoStop}%
\bibitem [{\citenamefont {Staub}(2014)}]{Staub:2013tta}%
  \BibitemOpen
  \bibfield  {author} {\bibinfo {author} {\bibfnamefont {F.}~\bibnamefont
  {Staub}},\ }\href {\doibase 10.1016/j.cpc.2014.02.018} {\bibfield  {journal}
  {\bibinfo  {journal} {Comput. Phys. Commun.}\ }\textbf {\bibinfo {volume}
  {185}},\ \bibinfo {pages} {1773} (\bibinfo {year} {2014})},\ \Eprint
  {http://arxiv.org/abs/1309.7223} {arXiv:1309.7223 [hep-ph]} \BibitemShut
  {NoStop}%
\bibitem [{\citenamefont {Porod}(2003)}]{Porod:2003um}%
  \BibitemOpen
  \bibfield  {author} {\bibinfo {author} {\bibfnamefont {W.}~\bibnamefont
  {Porod}},\ }\href {\doibase 10.1016/S0010-4655(03)00222-4} {\bibfield
  {journal} {\bibinfo  {journal} {Comput. Phys. Commun.}\ }\textbf {\bibinfo
  {volume} {153}},\ \bibinfo {pages} {275} (\bibinfo {year} {2003})},\ \Eprint
  {http://arxiv.org/abs/hep-ph/0301101} {arXiv:hep-ph/0301101} \BibitemShut
  {NoStop}%
\bibitem [{\citenamefont {Alguero}\ \emph {et~al.}(2024)\citenamefont
  {Alguero}, \citenamefont {Belanger}, \citenamefont {Boudjema}, \citenamefont
  {Chakraborti}, \citenamefont {Goudelis}, \citenamefont {Kraml}, \citenamefont
  {Mjallal},\ and\ \citenamefont {Pukhov}}]{Alguero:2023zol}%
  \BibitemOpen
  \bibfield  {author} {\bibinfo {author} {\bibfnamefont {G.}~\bibnamefont
  {Alguero}}, \bibinfo {author} {\bibfnamefont {G.}~\bibnamefont {Belanger}},
  \bibinfo {author} {\bibfnamefont {F.}~\bibnamefont {Boudjema}}, \bibinfo
  {author} {\bibfnamefont {S.}~\bibnamefont {Chakraborti}}, \bibinfo {author}
  {\bibfnamefont {A.}~\bibnamefont {Goudelis}}, \bibinfo {author}
  {\bibfnamefont {S.}~\bibnamefont {Kraml}}, \bibinfo {author} {\bibfnamefont
  {A.}~\bibnamefont {Mjallal}}, \ and\ \bibinfo {author} {\bibfnamefont
  {A.}~\bibnamefont {Pukhov}},\ }\href {\doibase 10.1016/j.cpc.2024.109133}
  {\bibfield  {journal} {\bibinfo  {journal} {Comput. Phys. Commun.}\ }\textbf
  {\bibinfo {volume} {299}},\ \bibinfo {pages} {109133} (\bibinfo {year}
  {2024})},\ \Eprint {http://arxiv.org/abs/2312.14894} {arXiv:2312.14894
  [hep-ph]} \BibitemShut {NoStop}%
\bibitem [{\citenamefont {Aalbers}\ \emph {et~al.}(2024)\citenamefont {Aalbers}
  \emph {et~al.}}]{LZ:2024zvo}%
  \BibitemOpen
  \bibfield  {author} {\bibinfo {author} {\bibfnamefont {J.}~\bibnamefont
  {Aalbers}} \emph {et~al.} (\bibinfo {collaboration} {LZ}),\ }\href@noop {} {\
   (\bibinfo {year} {2024})},\ \Eprint {http://arxiv.org/abs/2410.17036}
  {arXiv:2410.17036 [hep-ex]} \BibitemShut {NoStop}%
\end{thebibliography}%
\end{document}